\documentclass[twocolumn,aps]{revtex4}
\usepackage{graphicx}

\begin{document}
\title{KCQ: A New Approach to Quantum Cryptography \\
I. General Principles and Key Generation \footnote{The present
version v.6 is improved over v.4 which is identical to v.5.}}

\author{Horace P. Yuen}

\email{yuen@ece.northwestern.edu}
\affiliation{Department of Electrical and Computer Engineering,
  Department of Physics and Astronomy, Northwestern University,
  Evanston, IL 60208}

\begin{abstract}
A new approach to quantum cryptography to be called KCQ, keyed
communication in quantum noise, is developed on the basis of
quantum detection and communication theory for classical
information transmission. By the use of a shared secret key that
determines the quantum states generated for different data bit
sequences, the users may employ the corresponding optimum quantum
measurement to decode the data. This gives them a better error
performance than an attacker who does not know the key when she
makes her quantum measurement, and an overall generation of a
fresh key may be obtained from the resulting advantage. This
principle is illustrated in the operation of a concrete qubit
system A general information-theoretic description of the overall
approach will be presented, and contrasted with the
detection/coding description necessary for specific protocols. It
is shown that the attacker's error probability profile is needed
for a complete assessment of her information on the generated key.
The criterion of protocol efficiency and its sensitivity to system
parameter fluctuation is proposed as another benchmark on the
evaluation of key generation protocols. For systems described by
infinite-dimensional state spaces referred to as qumodes, KCQ key
generation schemes with coherent states of considerable energy
will be presented together with corresponding security analysis.
Various advantage enhancement and randomization techniques are
introduced for improving the security and efficiency of such
protocols.  A specific $m$-ary coherent orthogonal signaling
scheme, CPPM, is presented that can yield efficient secure key
generation over long-distance telecomm fibers using conventional
optical technology. The issue of secrecy in direct encryption
using KCQ is also discussed in general and in connection with the
$\alpha \eta$ protocol, on which experimental progress has been
made. It is indicated that information-theoretic security against
known-plaintext attack is possible, which has never been suggested
for any cryptosystem. In particular, it is shown that CPPM offers
information-theoretic security against known-plaintext attacks
while the data are unconditionally secure. Some qualitative
comparison among the different key generation schemes are made
from both a fundamental and a practical viewpoint. Further
quantitative development, the detailed analysis of direct
encryption, and the effects of various advantage enhancement
techniques would be presented in future papers of this series.
Some apparent gaps in the unconditional security proofs of
previous protocols are indicated in Appendix A. The core of the
paper is contained in sections III and VI.\\

\tableofcontents
\end{abstract}
\maketitle

\section{Introduction}\label{sec:intro}
Quantum cryptography, the study of cryptographic protocols with
security built on the basis of quantum effects, has been mainly
developed along the line of the original BB84 protocol \cite{bene}
and its variations \cite{gisi}. The focus is on {\it key
generation} (key expansion \footnote {The term `key expansion' is
often used in conventional cryptography to denote a (session) key
derived from a master key, which does not possess (perfect) {\it
forward secrecy} in the sense that its randomness is derived
entirely from that of the master key. Such a key is not fresh - it
does not possess randomness statistically independent of the
master key. We opt for the term `key generation' to signify that
the key generated is statistically independent of any secret key
used during the generation process.}), the establishment of a
fresh key \cite{mene} between two users, which is often referred
to in the literature as quantum key distribution \footnote {For a
comprehensive discussion of standard cryptography and its concepts
and terminology, some of which are transferred to the quantum
domain with a (confusingly) different meaning such as `key
distribution' and `key expansion', see Ref. \cite{mene}. We would
not use these terms in this paper, and instead use the term `key
generation' and {\it quantum key generation} (QKG) in their place.
With key generation, the usual `key agreement' in the strict sense
\cite{mene} can be obtained immediately. }. Without the use of
quantum effects, it was known that (classical) key generation is
possible whenever the user and the attacker have different
observations (ciphertexts) from which the user can derive a
performance advantage \cite{wyne,csis,maur}, a process to be
referred to as advantage creation \footnote{ This is more general
than the term `advantage distillation' used in the literature to
denote postdetection selection of the kind in Ref.
\cite{maur,ben2}. See Section III.B in the following. }.
 In BB84 type
\footnote {The other type based on  B92 \cite{yuen} is essentially
classical similar to that of Ref. \cite{ben2}, the idea of which
was already proposed in Ref. \cite{maur}. This is because
irreducible quantum noise in B92 may be replaced by irreducible
classical noise, and advantage creation (distillation) may be
obtained by post-detection selection that can be made on classical
noise systems. Its security is not compromised by the possession
of one copy of the observation by the attacker which is identical
to that of the user without a shared secret key. (This one copy
rather than full cloning capability is the proper analog to a
cloner in the classical situation, as explained in Section II.
Other points on this type of protocols would be elaborated
elsewhere.) It is merely a matter of terminology whether such
cryptosystems should be called `quantum'. Note that the
corresponding classical cryptosystem would also be unconditionally
secure if the noise cannot be ridden off by technological advance,
as for example, background sunlight. The main point is that in
such protocols advantage creation is not obtained from a quantum
effect that has no classical analog.} quantum cryptographic
schemes, advantage creation is obtained through intrusion-level
detection \footnote {We distinguish the qualitative {\it intrusion
detection} that tells the absence or presence of an attacker, from
the quantitative {\it intrusion-level detection} that is needed in
a BB84 protocol to generate a fresh final key that the attacker
knows essentially nothing about.} that quantitatively assures the
attacker's observation to be inferior to the users', thus allowing
{\it privacy distillation} (amplification) \cite{ben3} to
essentially eliminate the attacker's information on the final key
generated. Classically, this approach cannot succeed because the
attacker can always, in principle, clone a copy identical to the
user's observation, and no advantage can possibly be created.
Quantum mechanically, there is a general tradeoff between the
attacker's disturbance and her information on the user's
observation. By estimating the intrusion level, the user can
(probabilistically) assure a better observation for decoding the
original data, from which a fresh key may be generated.

There are several problems, in theory and in practice, with the
BB84 type \footnote{The classical noise and B92 type protocols
suffer from many of these problems also, due to  the necessary use
of weak signals. On the other hand, the post-detection selection
technique in these protocols can be used on top of KCQ protocols
to increase the security level at the expense of efficiency. }
quantum protocols. Among them are the necessity of using weak but
accurate signal source, a near perfect transmission line,
sensitive and fast quantum detectors, as well as the difficulties
of having appropriate amplifiers or repeaters to compensate loss,
developing specific practical protocols with quantfiable security
against all realistic attacks, and achieving reasonable efficiency
with such protocols. These problems are summarized in Section VIII
and a few hitherto un-analyzed problems are summarized in Appendix
A. As a consequence, the usefulness of BB84 type protocols is
severely curtailed, especially for commercial applications. Most
of these problems can be traced to the need of measuring the
intrusion level for balance between the user and attacker's
information on the data, and the necessity of using weak signals.
 In this paper, a new type of quantum protocols, to be called
KCQ (keyed communication or keyed CDMA in quantum noise),
 is presented. They do not
need to involve intrusion-level detection and permit the use of
coherent states with considerable energy, thus alleviating the
above problems. They can be implemented using optical technology
and readily integrated with existing optical communication
formats. It is hoped that they would quickly bring quantum
cryptography to practical application.

The basic idea of KCQ is to utilize a shared secret key between
the users to determine the quantum signal set \footnote{ The term
CDMA - code division multiple access - is used here as in cellular
communication to denote an arbitrary signal set for
communication.} to be chosen separately for each information
sequence, the quantum noise being inherent in the quantum signal
set from quantum detection and communication theory
\cite{hels,yue2,hole,yue3}. Such shared secret keys have also {\it
not} been used in classical key generation \cite{maur}.
 On the other hand, the use of a shared secret
key is necessary in BB84 type protocols and classical public
discussion protocols \cite{maur} for the purpose of
message authentication or realizing
the public channel. In contrast, a shared secret key is utilized in an
essential way on KCQ protocols, but a fresh key can be generated
that is much larger than the secret key used during key generation.
For KCQ key generation, advantage creation is obtained from
the different optimal or near-optimal
 quantum receiver performance between the user
who knows the key and the attacker who does not when
she makes her quantum measurement, even when a copy
of the quantum signal is granted to the attacker for the purpose of
bounding her information {\it without} intrusion-level detection.
This difference in performance has {\it no} classical analog. The
KCQ approach evolves from the anonymous
key encryption method described in Ref. \cite{yue4}.
 Exactly how and why
this approach works is explained both generally and concretely
in this paper.

\begin{figure*}
\includegraphics{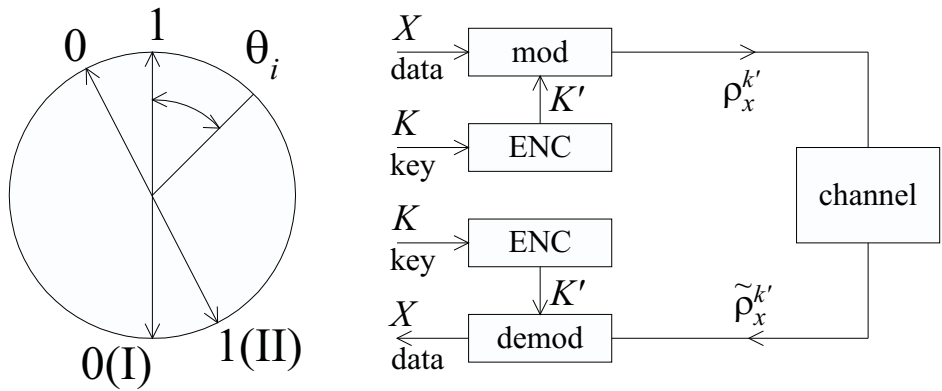}
\caption{The qk scheme. Left --- two neighboring bases, I and
 II. Right --- overall encryption involves modulation with bases
 determined by a running key $K'$ generated from the seed key $K$ via an encryption mechanism denoted by the box ENC.}
\end{figure*}

In practice, infinite dimensional state spaces to be referred as
{\it qumodes} provide the standard framework for describing
coherent-state laser signals. As in classical communication
systems, qumode systems allow the suppression of errors with
signal energy without error control coding that may complicate
security proofs and hinder the development of specific protocols.
Also, KCQ can be used for direct encryption, which has different
security performance criteria for key generation apart from the
inefficient one-time pad approach. The basic use of KCQ in binary
and $m$-ary detection of coherent-state qumode systems will be
described. In particular, a specific coherent pulse-position
modulation scheme, to be called CPPM, is shown to have many
dramatic characteristics including automatic privacy distillation,
secure key generation and data transmission over long-distance
telecomm fibers, and information-theoretic security against
known-plaintext attacks. This last characteristics is impossible
in conventional cryptography, the possibility of which has also
never been suggested in quantum cryptography.
 These qumode results are presented
in sections V and VI, and direct encryption briefly described section
VII. They are qualitatively compared to other quantum key
generation (QKG) schemes in section
VIII. Detailed quantitative development of CPPM and other schemes for
operation in realistic environment will be given in the future.

Key generation via KCQ on qubits is developed in Sections II. In
Section III, a general analysis of the KCQ approach and QKG
security against joint attacks are presented with the more
complete criterion of error profile instead of mere mutual
information.  A protocol efficiency criterion on QKG protocols is
introduced in Section IV, which should be insensitive to system
parameter fluctuation for a protocol to be realistically useful.
In Appendix A, we briefly discuss some serious gaps in the QKG
unconditional security proofs against joint attacks given in the
literature. In Appendix B, we respond to several criticisms on
$\alpha \eta$, a coherent-state KCQ scheme for direct encryption
upon which significant experimental progress has been made. For
readers who want to go directly to the core of our new results,
please read sections III, V, and VI.
\section{KCQ qubit key generation}\label{sec:qkg}
We consider a specific KCQ qubit scheme  for key generation, to be
called qk, to introduce and explain the characteristics of the KCQ
approach to quantum cryptography.

Let an arbitrary qubit state be represented by a real vector on
the Bloch-Poincare sphere. As depicted in Fig. 1, an even number
of $M$ points uniformly distributed on a fixed great circle on the
sphere, corresponding to $M/2$ possible orthonormal bases, are
used as possible quantum signal states for the bit value b $=0,1$.
The opposite points on a diameter of the circle for a given basis
are the two orthonormal states for the two possible bit values.
The two neighbors of each of the $M$ points are taken to represent
a different bit value. A shared secret key $K$ between two users,
Adam (A) and Babe (B), is used to select a specific basis for each
qubit. A secret {\it polarity bit} may also be introduced to be
added to the data bit for randomizing the polarity of the basis.
Instead of using the same $K$ for each b, a long running key
$K^{\prime}$ obtained from the output of a standard (classical)
encryption mechanism with $K$ as the input may be used to yield
different basis selection and polarity bits for different b's in
an $n$-sequence of input data $X_n$. This is depicted in Fig. 1
where, e.g., the ENC box may represent a synchronous stream cipher
or even just a linear feedback shift register (LFSR) for the key
extension \footnote{ We use the term `key extension' to denote the
process of getting a larger session key $K^{\prime}$ from a seed
key $K$, avoiding the term `key expansion' for possible
confusions.}. Thus, generally a total of $ \hspace{2mm}
1+\log_2(M/2) \hspace{2mm}$
 bits from $K^{\prime}$ would be used to determine
the polarity bit and the selection of one of $M/2$ possible bases.

The key generation process goes as follows. Adam picks a random
$n$-bit data sequence $X_n$, modulating $n$ corresponding qubits
by using $K^{\prime}$ to determine the polarity and basis for each
qubit. Babe generates the same $K^{\prime}$ to decide on the
quantum measurement basis for each b in $X_n$, and decode the bit
value by the corresponding measurement. A classical error
correcting code (CECC) may be used on the $n$-sequence to
eliminate noise in the system that may originate anywhere,
including source, transmission line, and detector. Privacy
distillation may then be employed to bring the attacker Eve (E)'s
bit error rate $P_E^b$ on the final key $K^g$ to any desired small
level. Intrusion-level detection is avoided by granting E a {\it
full} copy of the quantum signal for the purpose of bounding her
information. Advantage creation is obtained from the different
(optimal) quantum receiver performance, in an individual or joint
attack, between B who knows $K$ and E who does not. Further data
and signal {\it keyless randomization} may be introduced by A to
guarantee security against joint attack and to improve the key-bit
generation efficiency $k^g_{eff}$. Finally, the new key $K^g$ is
verified to be correct by the use of another short key $K_v$,
which may be done openly (publicly, i.e., the ciphertext is
available to E). All these steps and features would be described
and explained in detail in this and the following sections. There
are significant theoretical and practical advantage of the qk
scheme of Fig. 1 as compared to BB84. In particular, it can handle
all system imperfection and noise of any origin at the same time,
and produce reasonable key-bit generation efficiency $k^g_{eff}$.

The substantive use of secret keys in the key generation process
other than message authentication has been introduced before
\cite{yue4,hwan,barb}, In particular, it has been used in
\cite{hwan,barb} in selecting one of the two possible bases in
BB84 for improving $k^g_{eff}$, while keeping the other protocol
steps including intrusion-level detection intact. In contrast, the
security of our KCQ scheme is derived from a different quantum
principle, the difference in optimal quantum receiver performance
with and without the key, rather than the information/disturbance
tradeoff underlying BB84 and related schemes. Our use of a long
running key $K^{\prime}$ and large $M$ is not only essential in
obtaining high $k^g_{eff}$, it is also essential for obtaining
complexity-based security against known-plaintext attacks when KQ
is used for direct encryption. It also plays a role in yielding
reasonable key generation rates for coding-theoretic protocols
with unconditional security. In contrast, the unconditional
security proof described in Ref. \cite{barb} is not complete even
with intrusion-level detection, due to the quantum state
correlation or memory among different qubits induced by the shared
secret key, in addition to the previous problems in such security
proofs are described in Appendix A. Indeed, it is a major problem
in using a secret key that one needs to show there is a net
resulting key generated after subtracting the original key
 used, or that the system is somehow worth the $|K|$ cost.
Most significantly, the use of shared secret keys in our KCQ
approach makes possible the development of large-signal schemes
with conventional optical technology as described in sections V
and VI.

We first analyze the security of the above qk scheme under joint
attack on the seed key $K$ and a specific kind of individual
attack on the data $X$, given the quantum ciphertext. In contrast
to data encryption, there is no known-plaintext attack \cite{mene}
in key generation because one {\it presumes} A can generate
completely random data bits unknown to both B and E a priori. The
problem of how this can be done at high rate is a separate issue
common to every kind of cryptography. The term `individual attack'
is ambiguous, but the attack considered in the quantum
cryptography literature under this label usually refers to the
situation where E prepares the probe/interaction to each qubit of
the quantum signal sequence individually and {\it identically},
measures each of her probe individually and {\it identically}, and
processes the resulting information  {\it independently}
\footnote{We distinguish `independent' from `uncorrelated' in the
standard mathematical and statistical sense, avoiding the common
use of `uncorrelated' to mean `statistically independent' by many
physicists.} from one qubit to the other. Since, for bounding E's
information, we grant E a copy of the quantum signal sequence, the
optimal performance of which clearly provides an {\it upper bound}
on E's performance with an actual inferior copy obtained via a
probe, there is no question of probe/interaction in the attack on
such systems, only individual-qubit versus collective
measurements. Possible disruption of the signal by E will be
discussed in Sections III.F. The individual attack on the data
analyzed quantitatively in the following is of the same nature as
that in the BB84 literature \cite{gisi}, namely a constant
measurement on each qubit and independent processing. We will call
such attacks {\it constant individual attacks}.

Such an attack does {\it not} include all possible attacks within
a reasonable limitation on E's technology, in both the BB84 and
our schemes. If one may limit E's measurement to individual qubit
ones, perhaps because measurement across many qubits is difficult
to make \footnote{ Indeed, if there is already a good experimental
demonstration of a complete Bell measurement over a single pair of
single-photon qubits, there is none on 3-qubit systems.}, there is
no reason to limit E's classical processing after measurement to
qubit-by-qubit separately, except for ease of analysis! A more
detailed general classification of attacks and their analysis will
be provided in the future.

Generally, let $\rho_x^{k}$ be the quantum state corresponding
to the data $x$
\footnote{
Generally, we use capitals to denote a random variable and
lowercases to denote a specific value it takes, although the
distinction may not be necessary dependent on context.}
(single bit or a bit sequence) and running key
sequence $k$
\footnote{
We often use $k$ instead of $k^{\prime}$ for values of
$K^{\prime}$ to simplify notation, which should not cause confusion.}
that is used to determine the basis and/or polarity of
the qk scheme for that length of $x$. For $M/2$ possible bases
and a single bit $x$,
there is $\hspace{2mm}1+\log_2(M/2)\hspace{2mm}$ bits
 in $k$ for both basis and polarity
determination. Each $\rho_x^{k}$ can be represented as a real
vector $|r_x^{k} \rangle$ of norm $1$ on the great circle, the angle
between any two nearest neighbor vectors is $2\pi/M$ radian. The
quantum ciphertext available to E for upperbounding her
performance is $\rho_x^{k}$ where both $x$ and $k$ are
random.

For the purpose of attacking the key, the quantum ciphertext
reduces to $\rho^k= \sum_x p_x \rho_x^{k}$
where $p_x$ is the a priori probability of the data $x$. By
an optimal measurement on the qubits, the probability of
correctly identifying the key is obtained from $\rho^k$ via quantum
detection theory. We may let $x$ be any $n$-sequence, so that
$\rho_x^{k}=\rho_{x_1}^{k} \otimes \cdot \cdot \cdot
\otimes \rho_{x_n}^{k}$. Let $p_x= 1/2^n$ for each $n$-bit
sequence. It is easily seen from the quantum modulation format that
$\sum_x p_x \rho_x^{k}= \otimes^n (I/2)$ where $I$ is the
qubit identity operator, irrespective of the nature of $k$.
Thus, $\rho^k$ is independent of $k$ and the quantum ciphertext
provides no information on $k$ at all. Specifically, any processing on
$\rho^k$ yields an a posteriori probability on $k$ equal to the a priori
probability $p_k$, which may be chosen to be uniform for maximum
security.

Consider now the attack on $x$, through the decision on $x$
from measurement on the quantum ciphertext in state
 $\rho_x= \sum_k p_k \rho_x^{k}$. For individual attacks of
 the kind described above, one obtains $\rho_x$ corresponding
 to a single bit by tracing out the rest of the data sequence. When
 a polarity bit is used from $k$, it is easily seen that E's bit
 error probability is $P_E^b= 1/2$ by averaging on the two
 polarities alone. When the polarity bit is not used, $\rho_1$ and
$\rho_0$ are different. The optimum $P_E^b$ is given by
$(1/2)-(1/4)||\rho_1-\rho_0||_1$ \cite{hels}, in terms of the trace distance
$||\rho_1-\rho_0||_1$ between  $\rho_1$ and $\rho_0$, which can
be explicitly evaluated for a single qubit, with resulting
\begin{equation}
\label{a}
P_E^b=\frac{1}{2}-\frac{1}{M}
\left[ \frac{1-\cos(\pi/M)}{2\sin^2(\pi/M)} \right] ^{1/2}.
\end{equation}
This $P_E^b$ goes as $\frac{1}{2}-\frac{1}{2M}$ for large $M$
and can thus be made arbitrarily small.

That it is unreasonable to consider only such limited individual
attacks can be seen as follows. Suppose a key $k$ of $\hspace{2mm}
1+\log_2(M/2) \hspace{2mm}$ bits is used repeatedly to determine
the polarity and basis choice of each qubit state in a sequence.
Even though the above individual attack error rate for E is the
ideal 1/2, in actuality E has a probability $2^{-|k|}$, $|k|
\equiv $ number of bits in $k$, of completely decrypting $x$ by
guessing $k$ and using it on every qubit \footnote{ Note that this
possibility only obtains under our performance bounding assumption
that E has a whole copy, which would not occur in practice without
E disrupting the protocol so much that it would be aborted during
key verification - see section III.F and III.G. Nevertheless, it
demonstrates that it is sometimes possible for E to obtain a lot
more information by collective rather than independent processing.
}. In a similar way, the same problem arises for a running key
obtained from a short seed key $K$. That is, the correlation
between bits in $x$ due to $k$ can be exploited by joint classical
processing \footnote{ Similarly in BB84, joint classical
processing on individual qubit measurements may be used to exploit
the overall correlation between bits to optimize E's information
through the error correction information announced publicly. As to
be discussed further later, this has not been properly accounted
for in the security analysis in the literature for both individual
and joint attacks.} on individual qubit measurements. The presence
of the encryption box in Fig. 1 does not improve the situation
fundamentally for information-theoretic security even if the seed
key $K$ is long, because E can generate $K^{\prime}$ from a
guessed $K$ as the encryption mechanism is openly known \footnote{
This is the Kerckhoff's Principle in cryptography which states
that only the shared secret key can be assumed unknown to an
attacker.}.
Before we discuss general joint attacks in the next section, a number
of issues on the above development would first be cleared up.

First, aside from information-theoretic security issue, the
encryption box in Fig. 1 always increases the security of qk
through physical and computational complexity,
 and it also increases the efficiency of
key-bit generation fundamentally. Since there is often a
trade-off between security and efficiency,
 increasing the efficiency without compromising security is in
a sense increasing security (for a given efficiency). Suppose the
encryption box is a stream cipher that outputs a long running key
$K^{\prime}$ from a seed key $K$. If it is a maximum length LFSR
of $|K|$ stages, then $K^{\prime}$ up to length $2^{|K|}$ is
`random' in various sense, even though an exact knowledge of
$K^{\prime}$ of length $2|K|$ is sufficient to determine $K$
uniquely from the Berlekamp-Massey algorithm \cite{mene}. If $K$
is used repeatedly without $K^{\prime}$ in Fig. 1, the $\rho_x^k$
would be correlated by the repeated $k$ for short $x$-sequence,
with any reasonable $|K|$ and $M$. A joint (measurement) attack
can then be launched much more easily because the {\it physical
complexity} - in this case the correlated qubit measurement -
needed is much smaller than the one that comes from a long
$K^{\prime}$ from the same $K$. While $K^{\prime}$ is not open to
observation in the present case \footnote{ In some KCQ
implementations, in particular the ones on qumodes reported in
\cite{barb,corn},
 $K^{\prime}$ is open to partial observation.
Then direct complexity-based
 security obtains from the need to inverting
such imprecise $K^{\prime}$ to $K$.},
{\it computational complexity} obtains in any event
 when one attempts to
correlate the different bits in $x$ through the unknown key
\footnote{
Note that there is no need to know $k$ to be able to
correlate data bits through its repeated use. For example, in two
uses of one time pad $x_1\oplus k,x_2 \oplus k $
on two random data bits $x_1$ and $x_2$, we know
$x_1\oplus x_2 $ without knowing $k$. Indeed, no information on $k$
is obtained from the observation of
 $x_1\oplus k$ and $ x_2 \oplus k $.}.
Computational complexity is an excellent security mechanism if can
be shown to be exponential, as the Grover search can only reduce
the exponent by a factor of 1/2. Long keys with $|K|\sim 10^3$ can
readily be used in stream ciphers, and searching $2^{100}$ items
is already far beyond the capability of any imagined quantum
computer. In conventional cryptography \footnote{We use the term
`conventional cryptography' to denote the situation where E and B
have the same observation, $Y_E = Y_B =Y$. It is distinguished
from classical noise cryptography and from quantum cryptography.},
 many stream ciphers are used by
themselves as the complete security mechanism. It can be
incorporated in schemes such as Fig. 1 or the qumodes schemes of
Refs. \cite{corn,cove} to increase the overall security in direct
encryption.

Second, if cloning is possible so that $2^{|K|}$ copies of the
quantum ciphertext are available, there is no possibility of key
generation in principle. This is because E can use the $2^{|K|}$
different keys on the different copies, narrowing down the data to
exactly $2^{|K|}$ possibilities corresponding to the key
uncertainty. She can then follow whatever processing the users
employ on her own data, and the users have succeeded only in
obtaining a derived key, not a fresh one, whose randomness comes
entirely from the original key without forward secrecy. This {\it
also explains} why no key generation is possible when the user and
the attacker have the same observation.
 Note, however, the {\it difference}
between
cloning and having one full copy.  Having one copy is equivalent to
the classical situation where many identical copies can be made,
because together they do not tell the input data better than just one
copy. Quantum mechanically, the quantum uncertainty goes down with
the number of copies available from the laws of
quantum physics - indeed the state is in principle determined exactly,
say by quantum tomography, with an infinite supply of identical copies.
Thus, the classical analog of `cloning' is the granting of one identical
copy, not many ones as in full cloning.


Third, the possibility of attacking the data $x$ is not completely
described by $\rho_x$ in general. This is because E knows there is
a $\rho_x^k$ representation. Thus, she can attack the key via
quantum measurement on part of the qubit sequence, and then attack
the data with any knowledge she may thus learn. A similar but less
serious situation occurs for attacks on the key via partial
attacks on the data first. These possibilities, however, do not
arise in the case of constant individual attack.

Before turning to joint attacks, it should be noted that security
results against individual attacks are far from useless. In
practice, a joint attack may require correlated qubit measurement,
as the qubit states are correlated through the running key
$K^{\prime}$. Thus, the physical complexity of such quantum
measurement and the computational complexity introduced through
the encryption box would provide very significant security against
attacks that can be realistically launched in the foreseeable
future. When joint processing of the individual measurement
results is performed, the quantum noise introduced in such
individual measurement {\it already} yields a noisier copy for the
attacker that allows advantage distillation by the user. This
would distinguish our cryptosystem as a truly quantum one that has
no classical analog, and give real meaning to the security claim
against individual attacks. Performance under different types of
such `individual attacks', including ones involving adaptive
qubit-by-qubit measurements, will be presented in the future.

Consider the operation of qk of Fig. 1 for a perfect qubit channel,
under joint (collective) attack by E where, as a performance bounding
technique, she is supposed to have a full identical copy of the
n-sequence quantum state as B. She has to make a quantum
measurement, however, without knowing the key which she may possess
later --- see section III.B for a complete description. Here we would
observe that she would not be able to make a perfect decryption
with probability equal to $1$ for any finite $n$. After whatever
quantum measurement on the n-qubits she made, she still would not
be able to make a perfect decryption if K is then given to her.
This is because a perfect decryption occurs when and only when
the measurement she made is exactly that prescribed by K. This
shows that an advantage is created which may lead to an unconditionally
secure protocol with or without the use of further channel coding,
as shown in section III.H. When channel noise is present, a classical
error correcting code (CECC)
 on the quantum states may be employed
and separate privacy amplification may be required. The exact
quantitative performance will be detailed elsewhere.
\section{General principles of KCQ key generation}\label{sec:kcq}
In this section, the basic principles underlying key generation via
KCQ will be explained. The general principles of key generation will
first be reviewed and analyzed, extending the usual framework to
include
shared secret keys and the more appropriate criterion of error profile
in addition to and in place of mutual information. The quantum nature
of KCQ will be pinpointed. The overall steps and structure of a KCQ
key generation protocol will be exhibited, and the conditions for key
generation established. Various basic conditions on key generation
will be discussed, particularly in relation to the usual QKG approach.
This section III and the later qumode key generation section VI that
gives the specific qumode QKG protocol CPPM
 may be regarded as the heart
and brain of this paper. For a detailed review of the background in
classical as well as quantum detection and communication theory for
classical information transmission that is crucial for a complete
understanding of these sections, please see Ref. \cite{yue3} and
references cited therein. For a development of optical communication
theory that is important in fully comprehending the details of how
KCQ protocols work, please see also Ref. \cite{lo}.

Consider an entire joint process of
data transmission and encryption/decryption as described in Fig. 2.
A sends an $l$-bit sequence $U_l$ and encrypt/encode it into
an $n$-qubit or $n$-qumode
 sequence in state $\rho^k_x$ with the possible use of
a shared secret key $k$ with B,
which may include a source code key $K_s$, a channel code key $K_c$,
and a quantum state modulation code key $K_m$.
Classically, $\rho^k_x$ would be replaced by just an $n$-bit channel
input sequence $X_n$ corresponding to the $x$ in $\rho^k_x$.
The `channel'
represents all the interference from the system one has to suffer,
with $Ch^i$ giving output qubit states for $i=$ E, B. For E who does
not know $k$, the state is $\tilde{\rho}_x$ upon which she picks
a measurement on the basis of that and
her later knowledge from all sources
including public discussion to produce an estimate
 $\hat{K}^g_E$ of $K^g$, the final key generated by A and B.
 For B who
knows $k$, the channel output state is $\tilde{\rho}^k_x$ from which
she uses her knowledge of $k$ to obtain an estimate of
$\hat{U}^B_l$ of $U_l$. Classically, the states would be replaced by
 the
observations $Y^E_n$ and $Y^B_n$, the disturbed output of $X_n$.
Quantum mechanically, they are the results of corresponding
optimal or near-optimal measurements on the qubits or qumodes from
which the estimates $\hat{U}^B_l$ are made. One may first
consider, for simplicity, that $Y_n^E$ is obtained without
knowledge of $K_m$. More generally, one may split $\rho^{k}_x$
into parts from which attacks on $x$ and on $k$ are interwined.
Privacy distillation may already be incorporated in this process,
or may be added to $U_l$ and $\hat{U}^B_l$. The use of such an
approach for direct encryption is briefly treated in section VII.

\begin{figure*}
\includegraphics{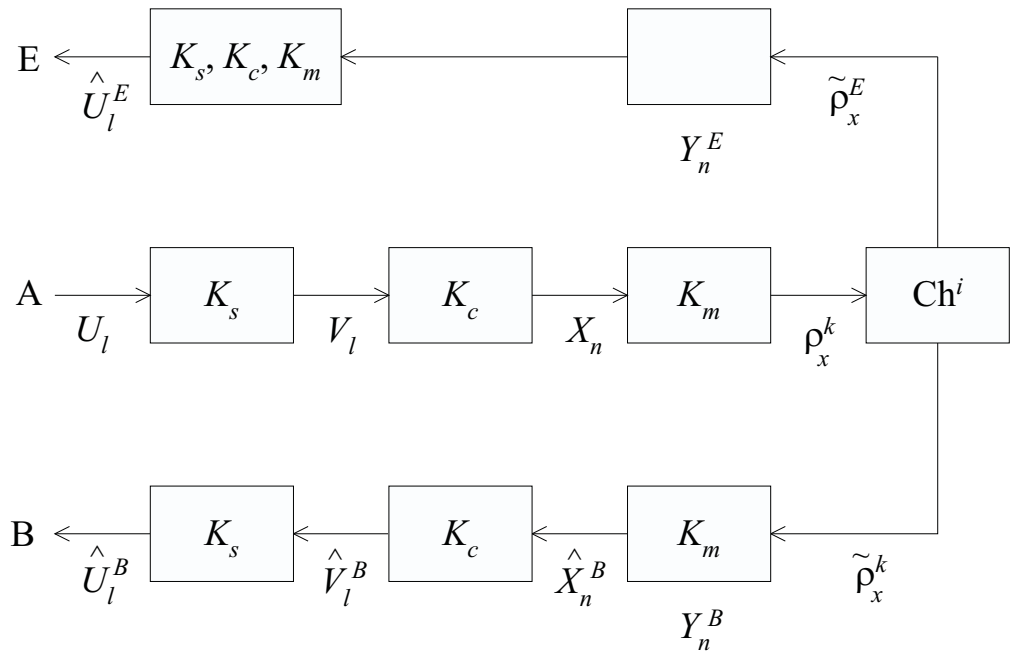}
\caption{General keyed communication in quantum noise.}
\end{figure*}
The essential steps in the operation of a KCQ key generation protocol
involve

\vspace{5mm}
(i) The use of a shared secret key $K_m$ between
A and B that determines the quantum states generated for the data
bit sequences in a detection/coding scheme between A and B that
gives them a better error performance over E who does not know
$K_m$ when she makes her quantum measurement;

(ii) A way for
A and B to extract a fresh key from the above performance
advantage;

(iii) A key verification process using another shared
secret key $K_v$ between A and B.
\vspace{5mm}

 The main novelty and power of
this approach, in principle, consists of

\vspace{5mm}
(a) Performance advantage is derived from the different quantum
receiver performance between B who knows the key $K_m$ when
she performs her quantum measurement and E who knows $K_m$ only
after she has made her quantum measurement.

(b) No intrusion-level detection or even intrusion detection is needed
by A or B.

(c) No public discussion is needed between A and B.

(d) No separate privacy distillation, or reduction in the key
generation rate due to any such equivalent operation, is needed in
a properly designed system. \vspace{5mm}

As a consequence, this approach makes possible the development of
an efficient, secure key generation protocol over long-distance
telecomm fibers using commercial optical technology. In the
following, these points will be fully explained and explicated.
The contrast between KCQ and the usual QKG approach, both in
theory and in practice, will be highlighted.
\subsection{Principle of Key Generation}
A key generation protocol with information-theoretic security,
whether it is based on classical or quantum randomness, would
consist of the following three logical steps:
\vspace{5mm}

(i) {\it Advantage Creation}:

The users A and B create a communication
situation between themselves with an observed random  variable
$Y_n^B$ for B that leads to a better error performance than that
obtained by E from her observed random variable $Y_n^E$ and all her
side information.

(ii) {\it Error Correction}:

The users agree on a generated string
that is free of error with high probability if E is absent.

(iii) {\it Privacy Distillation}:

The users derive from the generated
string a generated key $K^g$ on which E's
error probability profile satisfies a given security level.
\vspace{5mm}

The first step (i) may be achieved classically in the presence of
different noises for B and E's communication channels with respect
to A's data, using perhaps the help of public discussions between
A and B. In the quantum key generation approaches so far, (i) is
achieved via intrusion-level detection, explicit or implicit, that
guarantees that A and B have a better communication line than A
and E in the sense of mutual information, which is also the
privacy distillation criterion used in step (iii). The steps (ii)
and (iii) could be combined by an error correcting code, quantum
or classical, that simultaneously performs privacy distillation.
This is indeed the way the usual QKG unconditional security
(existence) proofs involving CSS codes \cite{lo2,gott,got2,lo3,
hama} are carried out. In QKG experiments to date, these steps are
distinct and a separate privacy distillation code is employed
whenever step (iii) is implemented. In our KCQ protocol called
CPPM in section VI, step (iii) is automatically achieved in an
ideal fashion from the $m$-ary signaling scheme employed. More
generally, we will show that privacy amplification is unnecessary
in most cases when the proper criterion of E's optimal error
probability is used in place of her mutual information.
\subsection{Advantage Creation with Shared Secret Key}
In the literatures \cite{wyne,csis,maur}, it was shown that if
a situation is obtained in which the mutual information between A and B,
$I(X_A;Y_B)$ for the random variables $X_A$ and $Y_B$ in
A and B's possessions, is bigger than that between A and E,
$I(X_A;Y_E)$, or that of B and E from the symmetrically
inter-changeable roles of A and B, key generation is possible. That
is, an information-theoretic existence proof is given under the
condition
\begin{equation}
\label{aaa}
I(X_A;Y_E) < I (X_A;Y_B)
\end{equation}
with the conclusion that an asymptotic key generation rate $\Delta
I =\max_{p(X_A)}[I(X_A;Y_B)-I(X_A;Y_E)]$ is possible between A and
B with the (total) amount of mutual information E has on the key
generated being arbitrarily small. In these results, there is no
shared secret key between A and B.

Such results can be generalized to include the use of a shared
secret key $K$ as follows. E is going to observe her channel
output $Y_E$ without the benefit of knowing $K$. However, one has
to make sure that the resulting $K^g$ generated between A and B is
fresh, i.e., statistically independent of $K$ and E's observation
$Y_E$, i.e. $I(K^g; Y_E K) \sim 0$. Indeed, E can try every
possible $2^{|K|}$ keys on her observation $Y_E$ to determine the
possible $X_A$'s. A conceptually convenient way to characterize
this situation is to give E the key $K$ after she made her
observation $Y_E$. Using the notation $I(X_A;Y_E K)$ to denote her
information in this situation where $Y_E K$ denotes the joint
random variables $Y_E$ and $K$, (\ref{aaa}) generalizes to
\begin{equation}
\label{aab} I(X_A;Y_E K) < I (X_A;Y_B)
\end{equation}
In (\ref{aab}), B is of course supposed to know $K$ when she plans
to observe $Y_B$. Classically, the condition (\ref{aab}) may
result if there is a limit on the data storage so that E cannot
have the same observation as B who can just store the relevant
data using the knowledge of $K$, as in the broadcast scheme of
Maurer. While the shared secret key $K$ is what occurs in the
above $I(X_A;Y_E K)$ with which E may use to estimate $K^g$ in KCQ
protocols, $K$ may be interpreted as all the side information E
obtains in other QKG protocols such as BB84, or as the missing
information that allows the conditional entropy $H(X|Y_E,K)=0$ in
a classical random channel protocol. These useful interpretation
would be used later.

The following important relation between any three random
variables should be noted:
\begin{equation}
\label{Z} I(X;Y K)= I(X;Y|K) + I(X;K)
\end{equation}
In the KCQ context, $I(X;K)=0$ and $I(X;Y_E|K)$ can be used in
lieu of $I(X;Y_E K)$. However, the distinction is important and
has various ramifications in the BB84 key generation when $K$ is
interpreted as $E$'s side information.

Quantum mechanically, the knowledge of $K_m$ that specifies the
mapping from classical data to quantum states would allow B to
choose the optimum or near-optimum quantum measurement to
discriminate among the data. Without knowing $K_m$, on the average
E would need to pick a quantum measurement that would allow her to
make reasonable estimates for different $k_m$'s, which leads to an
inferior performance compared to B for a specific $k_m$. This
situation clearly obtains when E does not have long-term quantum
memory to hold her copy of the quantum signal before she has to
make a quantum measurement to extract the information without
knowing the key. In practice, the key $K_m$ can be erased or kept
secret indefinitely from E, as E really would never have $K_m$,
and she has to make a quantum measurement without knowing $K$ {\it
even} if she has long-term quantum memory. Classically, there is
no need for E to know $K$ in order for her to be able to correlate
in a definite manner the different data connected by $K$, such as
the different session keys generated from a master key. For
example, with the observations of $x_1\oplus k$ and $x_2\oplus k$
for two independent bits $x_1$ and $x_2$ and a secret bit $k$, one
knows exactly $x_1 \oplus x_2$ while knowing nothing about $k$. In
this quantum situation, however, such correlation cannot be
obtained without quantum measurement on the quantum signals that E
possesses. Thus, E has to suffer the uncertainty of picking her
quantum measurement without knowing $K$ {\it even} if she has
long-term quantum memory. This is the {\it principle} underlying
advantage creation in KCQ protocols. Note that this is a quantum
effect with no classical analog, because classically E can always
make a complete observation of her received signal in principle.
There would be no incompatible measurements as in the quantum
case. The following intuitive result related to the advantage of a
shared secret key, as well as the average effect of side
information or missing information, is useful in the analysis of
classical and quantum key generation.

\vspace{5mm}
{\it Lemma 1}: For any three joint random variables $X,Y,K$,
\begin{equation}
\label{aac} I(X;YK) \leq I (X;Y) + H(K).
\end{equation}

{\it Proof}: From $I(X;Y K)= H(X, K) -H(X|Y,K)$,  (\ref{aac}) is
equivalent to $H(Y)+H(K)+H(X|Y,K) \geq H(Y|X) +H(X)$. The left
side of the last inequality is $\geq H(X,Y,K)= H(X)+ H(K|Y)+
H(X|Y,K) \geq H(X,Y) = H(X) +H(Y|X)$, completing the proof.
\vspace{5mm}

According to (\ref{aac}), if $\tilde{K}_m$ represents the missing
classical information in a channel with random parameter
 that together with $Y_E$ yields correct decryption,
or $H(X_A|Y_E,\tilde{K}_m)=0$, then $H(\tilde{K}_m)$ is indeed the
maximum possible amount of missed information, which may not be
large. For example, $\tilde{K}_m$ may represent the signal phase
or amplitude variation caused by a classically random channel. In
the quantum case, however, no missing information can make up
totally the loss from $Y_B$ to $Y_E$. There is irretrievable loss
from the inferior quantum measurement as the quantum copy is
expended upon measurement.
\subsection{Eve's Information and Error Profile}

The security criterion for key generation, classical or quantum,
has so far been limited to $I(K^g;Y^E_n)$, the mutual information
between Eve's observation $Y^E_n$ and the final key $K^g$
generated, with the provision that E's side information needs to
be accounted for \cite{niel}. Except in the limit $I(K^g;Y^E_n)=0$
which says that $K^g$ and $Y^E_n$ are statistically independent,
the information-theoretic quantity $I(K^g;Y^E_n)$ has no clear
quantitative operational significance with regard to the
usefulness of $Y^E_n$ for Eve in a eavesdropping context. One such
operational criterion is given by Eve's trial complexity $C_t$ as
measured by the average number of trials she needs to successfully
use trial keys on the basis of her information. For example, when
she knows nothing about $K^g$, which she would guess in successive
trials to, say, open a safe, she would need an average of $ C_t=
2^{|K_g|-1} + 1/2 \mbox{   } $ trials to succeed. It would also
describe E's ability when she launches a known-plaintext attack on
$K^g$ used in a standard cipher.
 In general, $C_t$ depends on her
exact {\it error profile \mbox{}} $p(\hat{K}^g_E= K^g|Y^E_n)$,
 the probability
that given her information, each of the $2^{|K_g|}$ guessed sequence
$\hat{K}^g_E$ is correct, as given in the following equation (\ref{A}).

This error profile E can obtain herself from the conditional probability
$p(X_A|Y_E)$, which is in turn specified by her channel transition
probability, the a priori data probability, and the overall
coding/communication format including possible deliberate randomization
by A. Furthermore, many different $p(\hat{K}^g_E=K^g|Y^E_n)$ leads
to the same $H(K^g|Y^E_n)$ or $I(K^g;Y^E_n)= |K^g|-H(K^g|Y^E_n) $,
which alone is not sufficient to capture E's ability to use her
information.
In the case  $I(K^g;Y^E_n)=0$, $p(\hat{K}^g_E=K^g|Y^E_n)= 2^{-|K^g|}$
for any $\hat{K}^g_E$. It is not known how $I(K^g:Y^E_n)$ is
related to $C_t$ in general except in the asymptotic limit when E
can encode or interpret $I(K^g;Y^E_n)$ via asymptotic equipartition
\cite{tree}, which she cannot since she does not control A's data
transmitter. One may
lower bound E's average bit-error
probability $P^E_b(K^g)$ by Fano's inequality \cite{tree},
which is valid for any n-sequence $Y^E_n$ that may possess
correlations among its bit values. Let $H_2$ be the binary entropy
function. Fano`s
inequality gives, in the present situation,
\begin{equation}
\label{aa}
H_2[P^E_b(K^g)] \geq 1- I(K^g; Y^E_n)/n.
\end{equation}
If $I(K^g; Y^E_n)$ is exponentially small with exponent $\Lambda$,
it follows from (\ref{aa}) that
$1/2- P^E_b(K^g)$ is exponentially small with exponent $\Lambda/2$.
However, the bit error rate is not really meaningful in this connection
because the bits may be highly correlated in the way they affect
$C_t$.

The following analysis shows that $I(K^g; Y^E_n)$ is not a
sufficient measure of E's capability unless it is really extremely
small. One needs to supplement it at least by $\bar{p}_E= \max
p(\hat{K}^g_E = K^g| Y^E_n)$, which is an important criterion in
our KCQ approach. The error profile gives the probabilities $p_1
\geq p_2 \cdot \cdot \cdot \geq p_N  $, $N= 2^{|K^g|}$, for each
of the $N$ guesses $\hat{K}^g_E$. Thus, $\bar{p}_E = p_1$ and the
trial complexity $C_t$ is
\begin{equation}
\label{A}
C_t = \sum_{n=1}^{N} n p_n.
\end{equation}
Given that E has $I_E$ bits of information on $K^g$, with $|K^g|=n$,
the largest $p_1$ that can be obtained is given by the error profile
that spreads $1-p_1$ among the $2^n-1$ other possibilities
uniformly, i.e. , it is determined by the equation
\begin{equation}
\label{G}
H_2(p_1) + (1-p_1) \log (2^n-1) = n- I_E.
\end{equation}
Thus, $p_1 \sim 2^{-l} $ for $I_E \sim n \cdot 2^{-l} $ and large
$n$, and E needs only about $1$ bit of information out of $|K^g|
=100$ for a possible error profile with $p_1=10^{-2}$, a
disastrous security bleach. This most favorable situation for E
with a given $I_E$ may be contrasted with her most unfavorable
situation, where her $1$ bit knowledge corresponds to knowing one
bit of the $|K^g|$-bit sequence exactly.

As we have just seen from (\ref{G}), $p_1$ can be made about as
large as the fraction of bits $|K^g|$ that she knows through $I_E$.
To ensure $p_1 \leq 2^{-l}$ through $I_E$,
 one needs to impose a strong
requirement that $I(K^g; Y^E_n) < n 2^{-l}$. In addition to the
necessity of assessing the actual minimum $C_t$ attainable with
a given $I(K^g;Y^E_n)$, the above condition on $p_1$ alone
would lead to $l \sim 100$ for a truly secure system, which is practically
very difficult to obtain on the basis of reducing $I(K^g;Y^E_n)$ for
realistic $n$.
On the other hand, a more detailed assessment of the error profile
or just $p_1$ would give a more accurate estimate of E's true ability
to use her information than the mere $I(K^g;Y^E_n)$ as shown in
the following.

Under the condition $p_1 \leq 2^{-l}$, one obtains

\vspace{5mm}
{\it Lemma 2}:

When E's optimum estimate of $K^g$ has an error probability
$\bar{P}^E_e \geq 1-2^{-l}$, her information on $K^g$ satisfies
$I(K^g;Y^E_n) \leq n-l, n= |K^g|$.

{\it Proof}:

Since $p_1 \leq 2^{-l}$, it follows that $l \leq n$. The maximum
$I_E$ is obtained when $H(\{ p_i\})$ is minimized, which occurs at
$p_1= \cdot \cdot \cdot = p_m = 2^{-l}$ for $m= 2^l$, $p_{m+1} =
\cdot \cdot \cdot = p_{2^n}= 0$, from the Schur-concavity
\cite{bhat}
 of $H$ (which can actually be seen to
follow from concavity directly).

\vspace{5mm}

Although this bound is weak, it may be used to establish the existence
of secure
KCQ protocols under rather general situations to be described in
section III.H.
Specifically, we take the approach that instead of $I_{E}$,
$p_E \leq 2^{-l}$ {\it has to be imposed}, which leads to
$C_t \geq (2^l +1)/2$ similar to Lemma 2.

\vspace{5mm}

{\it Lemma 3}:

Let E's optimal estimate of $K^g$ have a success probability
$\bar{p}_E \leq 2^{-\l}$, then her trial complexity $C_t$ is lower
bounded by $C_t \geq (2^l + 1)/2 $.

\vspace{5mm}

Thus, regardless of her mutual information $I_E$, Lemma 3
guarantees E's trial complexity at a level that may already be
satisfactory, and does that without privacy distillation. Note
that a bound on $C_t$ for given $I_E$ can be obtained via
(\ref{G}) and Lemma 3. It follows from Lemmas 2 and 3 that it is
much more useful to impose a bound on $p_1 \leq 2^{-80}$ for
$n=100$ that may correspond to $I_E \sim 20$, than a bound on $I_E
\leq 1$ that may corresponds to $p_1 \sim 10^{-2}$. Furthermore,
e.g., a system with
 $|K| \sim 20, n \sim 200, l \sim 100$ would be quite useful
 regardless of what $I_E$ actually is.

In view of our replacement of a constraint on $I_E$ by that of
$\bar{p}_E$ together with the corresponding elimination of privacy
distillation, the following comment is in order. Privacy
distillation moves E's uncertainty among $n$ bits to $n-l$ bits
when $I_E \sim l$ bits. If $K^g$ is to be used as one-time pad,
such a move is necessary for near perfect security level. However,
$K^g$ would often be used as a seed key in some other cipher. In
that case the privacy distillation merely reduces the key size of
$K^g$, not its randomness to E as already observed in Ref.
\cite{ben2}. Particularly for stream ciphers which may have
similar speed and complexity for different key sizes, it may not
be worth the trouble to carry out any privacy distillation before
the raw $K^g$ is used.

In attacking a KCQ system, E may always guess the seed key $K_m$
and make corresponding optimum quantum measurements to decrypt the
data. Her success probability, however, is only $\bar{p}_E=
2^{-|K|}$. If she guessed the key $K_m$ incorrectly, the
probability she would get $K_g$ correctly from $Y^E_n$ is
exponentially small in $n$ because different keys $k_m$ lead to
different $\rho^k_x$ for the same data. Furthermore, she has only
(at most) one copy of the quantum signal (with energy at the
designed security level in the qumode case) to launch this attack
once. This corresponds to the above error profile with $p_1=
2^{-|K_m|}$ but no more than one trial. Note that the seed key
$K_m$ has to be guessed in total as it is not being used in a bit
by bit or segment by segment manner.
\subsection{Key Rate of Secure Protocols}
As is evident form Fig. 2, E has to deal with the $X_n$ chosen by
A and has no independent way to encode her own channel
$I(X_n;Y^E_n)$. Thus, by choosing a rate $R$ in between
(\ref{aaa}) and (\ref{aab}), one may be able to force the second
term in $\Delta I$ to be zero and obtain $\Delta I = \max_{p(X_n)}
I(X_n;Y^B_n)$ as a consequence of the Shannon Coding Theorem and
its Strong Converses \cite{tree,cald,gros,yue8} for memoryless
channels. The Strong Converse states that the block error rate
goes to $1$ exponentially in the block length \cite{csis,han} at
rates above capacity,
 which may already imply
$I(X_n;Y^E_n)\rightarrow 0$ as will be seen in the CPPM scheme of
section VI. When a shared secret key is used, the classical channels
become ones with memory, and the corresponding Coding Theorems and
their Converse \cite{cald,gros,yue8} may be employed. Since such
results on channels with memory are typically weaker quantitatively,
we may employ the memoryless results using the following approach.
A may transmit at an input bit rate $R$ satisfying
\begin{equation}
\label{aad} I(X_n;Y^E_n K)/n < R < I(X_n;Y^B_n)/n
\end{equation}
where $I(X_n;Y_n)$ refers to the mutual information between
$n$-bit sequences. Let independent keys $K_i$, $i \in \{ 1,..., m
\}$, be the secret keys used for the $i$th $n$-sequence $x^i_n$,
each randomly chosen from $\{0,1\}^{|K|}$. Then, by treating each
$x_n$ sequence as a single word or symbol, we have created a
memoryless channel on the symbols. This is because both E and B's
channel transition probabilities for the different $m$ symbols are
the same, respectively. As a consequence, the memoryless Coding
Theorem and its Strong Converse can be applied to yield the
existence proof of a code that generates arbitrarily close to
$mI(X_n;Y^B_n)$ bits.

It should be noted that condition (\ref{aad}) already includes the
cost of the key $|K|$ for net key generation. If one uses instead
the condition
\begin{equation}
\label{Y} I(X_n;Y^E_n)/n < R < I(X_n;Y^B_n)/n
\end{equation}
together with the net-key generation condition
\begin{equation}
\label{X} |K| < \Delta I \equiv I(X_n;Y^E_n)- I(X_n;Y^B_n),
\end{equation}
it is more stringent than (\ref{aad}) from Lemma 1. If the system
is information-theoretic secure against known-plaintext attacks
for direct encryption, briefly discussed in section VII and
treated extensively in Part II, the keys may be re-used because
the different fresh keys generated from two different uses of $K$
are independent of each other. In such a case, only condition
(\ref{Y}) without condition (\ref{X}) needs to be satisfied. If
the $K_i$ are not re-used, a net key generation rate is obtained
under (\ref{aab}), or (\ref{Y}) and (\ref{X}).

More generally, since the above strategy yields a block error rate
$P^E_e > 1- e^{-m \varepsilon}$ where $\varepsilon$ is a
characteristic of the channel that depends on $n$, and we have
 $\bar{p}_E = p_1 \leq e^{-m \varepsilon}$ even when the error-free
 data is used as $K^g$ without additional privacy distillation.
 Using lemma 2, we have also
$I(K^g; Y^E_{nm}|K) \leq m(n- \varepsilon)$. To summarize, using
also lemma 3 we have

\vspace{5mm}
{\it Theorem 1}:

Under condition (\ref{aad}), unconditionally secure protocols may
be obtained via error correction coding but without further
privacy distillation that satisfy
\begin{equation}
\label{B} \bar{p}_1 \leq e^{-m \varepsilon}, \mbox{           }
I(X_n; Y^E_{nm}|K) \leq m(n- \varepsilon), \mbox{           } C_t
\geq (2^{m \varepsilon} +1)/2.
\end{equation}
where $\varepsilon >0$ is determined from the channel specification.
\vspace{5mm}

Note that a net key $K^g$ can be taken as $X_n$ or its privacy
distilled version with $I(K^g; Y^E_{nm} K) \rightarrow 0 $
guaranteed from (\ref{B}) only when $|K| < \varepsilon$. On the
other hand, the security level (\ref{B}) itself may be
satisfactory already. We use the phrase `unconditional security'
above in the usual sense,
 except that the security
level is measured by $\bar{p}_E$ and $C_t$ rather than $I_E$. This
represents a more efficient approach because, as we have seen in
III.C, a bound on $\bar{p}_E$ has to be imposed in any case.
\subsection{Specific Detection or Coding Scheme}
As in all the unconditional security proofs of QKG protocols so
far presented in the literature, the above development can only
yield an existence proof of a secure protocol, because no specific
code has been given. For actual application, one would need to
provide a specific coding or signaling scheme, in either classical
or quantum key generation, and show quantitatively that Eve has
little knowledge on $K^g$. While $I(K^g;Y^E_n)$, to be
supplemented by E's side information, is a measure on E's
information, it is not the most useful quantity to deal with as we
have already seen above because E cannot encode. Indeed, as
discussed in III.C, it is her error profile on $K^g$ obtained from
$Y^E_n$, not just $I(K^g;Y^E_n)$,
 that really matters. The appropriate
measure to this end is her optimum block error probability
$\bar{P}^E_e(X_n|Y^E_n,K)$ given her observation and side information.
 For advantage creation, one wants to obtain the situation where
\begin{equation}
\label{aae} \bar{P}^E_e(X_n|Y^E_n,K) \rightarrow 1, \mbox{      }
 P^B_e(X_n|Y^B_n) \rightarrow 0.
\end{equation}
Condition (\ref{aae}) is equivalent to the above coding-theoretic
existence result when one argues, as in  (\ref{aad}), by coding on
the $n$-bit symbols. In actual application, it may well occur that
the users work with only a single $n$-bit symbol at a time,
especially that already corresponds to $n$-bit coding. In such
case,
 (\ref{aab}) or (\ref{aad}) can be ignored and   (\ref{aae}) is
 {\it the appropriate condition for advantage creation}.
Similar to Theorem 1, we obtain via Lemma 2 and Lemma 3.

\vspace{5mm}
{\it Theorem 2}:

For a detection scheme that has $\bar{P}^E_e(X_n|Y^E_n, K) > 1-
2^{-l}$ while $P^B_e(X_n|Y^E_n)$ is error-free in the use, an
unconditionally secure protocol for the following fixed security
level is obtained without further privacy distillation and with a
key cost $|K|$,
\begin{equation}
\bar{p}_E \leq 2^{-l},  \mbox{      }
 I(X_n;Y^E_n|K) \leq n-l, \mbox{      }
 C_t \geq (2^l + 1)/2.
\end{equation}
\vspace{5mm}

Note that Theorem 2 is valid for a single use of the block
detection scheme without further coding. As in Theorem 1, the
scheme is certainly useful whenever $|K|<l$ if the resulting
security level is satisfactory. To decrease $I(K^g; Y^E_n K)$,
further privacy distillation may be needed, while no guarantee is
offered in the theorem that a net key with $I_E \rightarrow 0$ can
be obtained that is larger than $|K|$. However, in
 an appropriately designed scheme, it may already happen that
 $I(X_n;Y^E_n K) \rightarrow 0$ with a secure error profile without
 further privacy distillation.
In such a case, the following bit-error rate condition holds with
independent bit errors,
\begin{equation}
\label{aaf} P^E_b(X|Y^{E}, K) \rightarrow \frac{1}{2}, \mbox{ }
P^B_b(X|Y^B) \rightarrow 0.
\end{equation}
See section VI for an example of a specific protocol that displays
such behavior under rather general attacks.
 Condition (\ref{aaf}) implies (\ref{aae}), and can be
taken as the advantage creation condition at the bit level that
requires no further privacy distillation for generating $K^g$.

Observe that in the security proof of a specific detection/coding
scheme, one must be careful to ascertain E's optimum block error
by including her adaptive attacks, in particular ones via attacks
on the key. Also, for a full security proof one needs to solve the
{\it novel}
 quantum
detection problem in which one chooses the optimum quantum
measurement in anticipation of making a future decision on the
basis of further information not available at the time of quantum
measurement. However, such problems {\it also} occur in the usual
QKG protocols even in the context of individual attacks, whenever
a specific scheme is employed rather than a mere coding-theoretic
existence claim. They have yet to be dealt with in the literature.
\subsection{Key Verification}
A final {\it key verification} process is needed in KCQ protocols as
compared to BB84. In this case, A or B uses the generated key
$k^g$ to encrypt a fixed shared secret bit sequence
$K^{\prime}_v$, which
is perhaps the extended output of a fixed known transformation on
some separate shorter shared secret key $K_v$,
$|K_v| \leq |K^{\prime}_v|$,
 and
sends it to the other party through reliable communication.
 The encryption mechanism for getting $K_v^{\prime}$ via $K^g$
may be the same as Fig.2 but used for direct encryption, thus no
privacy distillation and key verification would be involved there.
It is also possible to reverse the above roles of $K_v$ and $K^g$
in the verification, making it similar to a message authentication
protocol on $K^g$ with a secret key $K_v$. When  $K_v$ is used as
one-time pad to check an openly chosen random unkeyed hashed
version of $K^g$ of size $|K_v|$, the average probability that two
different $K^g$ are mistakenly agreed upon is $2^{-|K_v|}$. This
results from a random coding argument similar to the proof of the
Shannon Channel Coding Theorem or the privacy distillation code
performance theorem in Ref. \cite{ben3}. The standard results on
hashing collision \cite{mene} can also be used instead.
 If the users believe there is only a small
number of errors in $k^g$, they may try to correct them via open
discussion as in some BB84 protocols, via parity check or other
methods.

Given that a common key $k^g$ is generated between A and B, it can
be seen that E's information cannot be more
 than what
 she can get from one full copy of the quantum ciphertext,
 which may be granted to her for the purpose of bounding her
 information.
 This is because whatever probe she introduced that may mess
 up the state B receives, she cannot obtain more information than
that of a full copy although she may introduce enough errors to make
$K^g$ agreement between A and B impossible.
Such a mess up, however, is not something A and B could avoid in
the presence of even a `passive' attacker that takes energy out of the
signal by tapping. Thus, there is no loss of generality in not being
able to establish a key in the presence of E, so long as the protocol
is not sensitive as to be discussed in section IV. Also, E cannot
correlate her own states and B's states to obtain information, as she
does not know $K$ and what states to correlate with.
The probability of such successful attack is small exponentially in the
number of bits $n$. Thus, perfect
 forward secrecy of $K^g$ with respect to $K$ is obtained,
 while both intrusion detection and intrusion-level detection are
 avoided.
\subsection{Advantage Enhancement}
Given that advantage can be created, it is possible to {\it enhance} it,
i.e., decreasing the attacker's performance for a fixed user
performance, by various techniques. These include
first of all {\it data bit randomization} (DBR), the use of a randomly
chosen open or secret source code that re-arranges the $U_l$
in Fig. 2. Secondly, one may employ {\it deliberate error
randomization} (DER), the addition of error bits to B introduced
deliberately by A with a corresponding error correcting mechanism
such as a  CECC
 on the quantum states. Thirdly, one may introduce
{\it chaining} among the data and keys, i.e., the use of local feedback to
make future transmissions dependent on past ones. Especially when
used in conjunction, such techniques could lead to a flattening of
E's error profile for any fixed $n$, and hence her trial complexity
$C_t$ whether her entropy $H(X_n|Y^E_n,K)$ is affected or not.

In the classical situation where a fixed amount of missing information
represented by $\tilde{K}_m$ can be used to uniquely
 specify  the channel suffered by E as described in III. B,
all such techniques cannot produce $H(X_n|Y^E_n, K)$ more than
$H(\tilde{K}_m)$ from lemma 1. Nevertheless, such technique may still be
useful in a given application because $|\tilde{K}_m|$ may
be large or it may
be difficult to ascertain exactly. Furthermore, this is not the amount
needed to be provided as shared secrets between A and B. Quantum
mechanically, there is no $\tilde{K}_m$ that can restore the data perfectly
for E in a KCQ protocol. Thus, these techniques could increase the
key generation rate beyond the original $\Delta I$. In both the
classical and quantum cases, such possibility arises because the
joint probability distribution of the
random variables $(X_A, Y_B, Y_E)$ are not specified a priori,
but rather subject to the creation of communication lines
between (A, B) and (A, E) in a given communication situation.

Another technique for enhancing the advantage may be obtained as
follows. Let $1-\lambda$ be the fraction of system splitted off by
E and $\lambda$ the one remaining for B. This can be easily
quantified in qumode system via the signal energy, so that
 $\eta \lambda$ is the total fraction received by B under
 the line
 transmittance $\eta$. Let $p^B_{\eta \lambda}$ be the probability
that A and B verify that their generated keys agree with a fraction
$\eta \lambda$ of the signal received by B, and $p^E_{1-\lambda}$
the probability that E correctly obtains the key with her fraction
$1-\lambda$. The strategy of granting E a copy of the quantum
signal to bound her information is equivalent to the
condition that  E's probability of successful cheating $p^s_E$,
$p^s_E = p^B_{\eta} \cdot p^E_1$, is small
when A and B proceeds as if E were not interrupting
 in a KCQ protocol described in III. F.
By making $p^B_{\eta\lambda}$ small for
$ \lambda \leq \lambda_0 $, a set threshold, $p^S_E$ is modified to
\begin{equation}
\label{aag}
p^S_E = p^B_{\eta \lambda_0} \cdot p^E_{1-\lambda_0}.
\end{equation}
This represents advantage enhancement since (\ref{aag}) is smaller
than $p^B_{\eta} \cdot p^E_1$.

Two remarks on this technique are in order. First, under the use of
 (\ref{aag}) which is a kind of pre-set automatic intrusion detection,
 the cryptosystem becomes more sensitive and thus
loses some of the robustness characteristics of KCQ protocols.
Second, this technique may be considered as one of advantage
creation, because the user's performance may be correspondingly
lowered with the decrease of the attacker's performance.
\subsection{Overall KCQ Protocol}
Schematically, a KCQ protocol corresponding to the communication
situation of Fig. 2 may be summarized as follows.

\vspace{5mm}
{\it Generic KCQ Protocol:}

(i) A picks a random bit sequence $U_l$, encodes and modulates the
corresponding $n$ qubits or $n$ qumodes as in Fig. 2, with a total
secret key $K= (K_s, K_c, K_m, K_v)$ shared with B.

(ii) From $K_m$, advantage creation is achieved via the different error
performance obtainable by B and E who does and does not know
$K_m$ at the time of their quantum measurements.

(iii) Advantage enhancement and privacy distillation may be
achieved with appropriate system design, deliberate randomization
and chaining techniques, so that a substantial net key can be
generated on which E has as little information or as large an
optimum error probability as desired.

(iv) A and B verify that they agree on a common $K^g$ by using it with
the secret key string $K_v$.
\vspace{5mm}

It is important to note the crucial role of the key verification
process in the overall protocol. If E messed up the state B
receives, it produces {\it no} effect on the security of the
protocol if the key is verified because E {\it cannot} in any case
have more information on the correctly agreed key than what she
can get from one full copy of the quantum signal, exponentially
probabilistically. If the system is designed to be not sensitive
to small disturbance, as any practical system must be, it is
perfectly fine that the presence of E would disrupt the key
generation process so that the key is not verified. On the other
hand, there are problems in QKG protocols with intrusion-level
detection not to be elaborated on section IV.

We have explained the above steps that may enhance the advantage
and improve the efficiency and security of the protocol. A full
treatment of specific schemes will be given in the following and
in Part II. By combing the analysis of sections II and III, we
have

\vspace{5mm}
{\it Theorem 3}:

In the absence of channel noise, the protocol $qk$ of Fig. 1
allows unconditionally secure key generation for any fixed
$n$-sequence with a security level given in the form (\ref{B})
without further privacy distillation. \vspace{5mm}

{\it Proof}: We have seen in section II that E's error probability is bounded
away from zero for any $n$, while B's is exactly zero, and  the key
$K$ is completely hidden with the quantum ciphertext alone. Thus, the
above generic KCQ protocol would generate a fresh key.
By coding as in Theorem 1, the security level (\ref{B}) is obtained
where $\varepsilon$ depends on the encryption mechanism and the
state $\rho^k_x$.

\vspace{5mm}

Intuitively, one may expect that $H(K|X_n, Y^E_n)$ would remain
substantial in $qk$ even for large $n$, and also that
 $H(X_n|Y^E_n, K)$ is large. However, in the absence of either
 a security proof against known-plaintext attack or a proof that
 $I(X_n;Y^E_n K)/n$ can be made sufficiently small, Theorem 3 is not
 sufficient to guarantee a nonzero net key generation rate that E
 knows essentially nothing about.
On the other hand, the bounds on $\bar{P}_E$ and $C_t$ should be
sufficient. A similar result would hold in the presence of channel noise,
but a rigorous proof requires new techniques to be presented in
Part II.

 More generally, secure protocols can be created by using quantum
 entanglement as follows. For each $n$-bit data sequence $x$, let
 $2^{|K|}$ mutually orthogonal states in
 $\otimes^n_{i=1} H_i$ be the possible $\rho^k_x$ corresponding to
 different values of $K$, $|K| <n$. Each individual state space $H_i$
 may be of any dimension. There exist many such modulation formats
 where the resulting $\rho_x = \sum_k \rho^k_x$ are not mutually
orthogonal for different $x$. We have

 \vspace{5mm}
{\it Theorem 4}:

In the absence of channel noise, a KCQ protocol employing the
above state modulation allows unconditionally secure key
generation for any fixed $n$-sequence with security level given by
(\ref{B}) without further privacy distillation.

\vspace{5mm}
{\it Proof}:

From Lemma 1, $I(X_n;Y_n|K)$ as obtained by optimizing over all
E's possible quantum measurements is bound by $H(K)+I(X_n;Y_n) $,
obtained by the same measurement. From Holevo's inequality
\cite{niel}, $I(X_n;Y_n) \leq S(\sum_x p_x \rho_x) -\sum_x p_x
S(\rho_x) $. From the above modulation format, $S(\sum_x p_x
\rho_x)<n $ while $S(\rho) = |K| =H(K)$. Thus, condition
(\ref{aab}) is satisfied and unconditionally secure protocols
exists, via coding $n$-bit blocks described in III.D, with a
security level given by (\ref{B}). \vspace{5mm}

For both Theorems 3 and 4, the system could be useful for any
$|K|< n$ while
 the $|K_v|$ cost is negligible for large
$m$, but the net key generation rate is not guaranteed to be
nonzero if $I(K^g;Y_n^E)$ has to be made extremely small. Note
that the possibility of security proof for QKG relies on the claim
that all the possible useful actions of E have been exhausted.
That this is true in any particular protocol has to be ascertained
carefully. We assert that this is the case for KCQ protocols
described in this section and section VI,
 as long as the model is taken to be a valid description of
the real situation \footnote{ In general, one may ignore
insecurity claims against the security of a protocol that are made
for reasons not intrinsic to the protocol, e.g., that a shared
secret key is not really secret. Such a claim is common to all
protocols, which always require some common secrecy between two
users, say for agent authentication, that distinguishes them from
other parties. Similarly, the record of a secret key in KCQ
schemes can be assumed safeguarded or `destroyed', as the
situation is different from that of a public key distribution
center which needs to use the public key repeatedly. }.

Note that although security against attacks on the key is
automatic in qk, it is not in a general KCQ system. Even with key
security on quantum ciphertext-only attacks, one needs to consider
situations where some knowledge on the data $x$ is obtained from a
partial attack on $\rho^{k}_x$, and then an attack on the key $k$
is launched with such knowledge, and then on $x$ again, etc. The
system is not fully secure unless the key is secure against such
`statistical attack'. This problem also occurs in BB84 type QKG
when the generated key $K^g$ is used later for direct encryption.
Even for the one-time pad mode, a known-plaintext attack can be
launched to learn part of $K^g$, and knowledge on the rest may be
obtained by E from this and her probe information. These problems
will be analyzed in detail in part II.
\subsection{Unconditional Security and System Implementation}
Unconditional security (US) in QKG is usually taken to mean
security against all possible attacks allowed by the laws of
physics (and logic), at a level that can be made arbitrarily close
to perfect. In particular, it would imply security against an
attacker that has unlimited computational power and hence can
perform any exhaustive search. Unconditional security must
therefore be information-theoretic,
 not complexity-based,
security. The term has, lately, often been used in a weakened sense,
such as `unconditional security against individual attacks'. As long as
the security claim is precisely spelled out, the terminology issue is
secondary. Since the security is only as good as the mathematical
model being valid at the time of system use, it is useful to
consider various different qualitative degrees of unconditional
security, especially in commercial type applications. As a matter
of fact, the experimental development of QKG is still struggling
 within the
realm of individual attacks, the justification being that it is physically
complex, and currently practically impossible,
to launch more general attacks.
In this paper, various weaker claims of security are also considered,
especially for specific quantifiable protocols.

In the presence of system imperfection including those arising from
the source, transmission line, and detector, we separate out E's
disturbance and call the rest `channel noise' and `channel loss', as is
common in communication theory. In a situation intended for
cryptographic applications, one first determines what this actual
{\it channel} including all these imperfections is, as characterized by the
channel parameters in a canonical representation with some
confidence interval estimates on these parameters. To guarantee
security, advantage creation is to be obtained under the following
Advantage Creation Principle for unconditional security.

\vspace{5mm}
{\it (US) Advantage Creation Principle:}

\vspace{3mm}
All the noise and loss suffered by the users are assumed absent to
the attacker, except those arising from a fundamentally inescapable
limit or introduced deliberately by the user at the transmitter.
\vspace{5mm}

Generally, advantage creation is obtained from the difference
between $I(X_n, Y^B_n)$, which includes all the channel
disturbance not counting E's for KCQ schemes and which is obtained
from $\rho^{AB}$ after disturbance by E  for BB84 type schemes,
and $I(X_n; Y^E_n K)$ which includes all the irremovable system
disturbance
 to E.
There should be a sufficient margin between $I(X_n;Y^B_n)$ and
$R$ of (\ref{aad}) in a coding based
 KCQ scheme so that the protocol is not sensitive to small
fluctuation in channel parameters and small disturbance by E. Thus,
it is evident that even if one can in principle get $R$ close to
$I(X_n;Y^B_n)$,
 $\Delta I$ determines
the levels of channel parameter
fluctuation and E's disturbance that can be
tolerated for an efficient protocol (high $P_{eff}$ in Section IV).
 What counts as an irremovable disturbance to E is a matter of
technology state and reality constraints, as well as fundamental
obstacles. For unconditional security, such fundamental obstacles may
include the laws of physics, deliberate actions by A at the transmitter,
shared secret between A and B, as well as facts of nature as we
know them. Only these should be included in the above US Advantage
Creation Principle. For many applications, practical advantage
creation such as those obtained from
security against individual attacks, may be
quite sufficient.
Thus, in a weakened form, one may modify the US Advantage
Creation Principle by imposing on E whatever constraint that may
seem reasonable in a given application.
\section{Performance efficiency of QKG protocols}\label{sec:eff}
The performance of a QKG or key generation scheme for useful
real-life application is gauged not only by its security level, but also
its efficiency in at least two senses to be elaborated in the following.
The security level is most usefully measured in terms of
 E's error profile on the final key
as averaged  over the random parameters
 of the system and minimized over E's
possible attacks. Since there is generally a trade-off between
security and efficiency, raising the efficiency for a given security level
is equivalent to raising the security level for a given efficiency.
In addition, for a protocol to be useful the efficiencies cannot be
too low.

The first type of efficiency that should be considered is {\it protocol
efficiency}, denoted by $P_{eff}$, which has not been treated in the QKG
literature. It can be defined as the probability that the protocol is not
aborted for a given channel and a fixed security level
in the absence of an attacker E. It is essential to consider the
robustness of $P_{eff}$ with respect to channel parameter fluctuation,
e.g., how {\it sensitive} $P_{eff}$ is to small changes in channel
parameter $\lambda_c$ which may denote, e.g., the independent
qubit noise rate of any kind. In practice, $\lambda_c$ is known only
approximately for a variety of reasons, and imperfection in the system
can never be entirely eliminated. If $P_{eff}$ is sensitive to such small
changes, the protocol may be practically useless as it may be
aborted almost all the time. Sensitivity issues are crucial in engineering
design, and there are examples of `supersensitive' ideal system
whose performance drops dramatically in the presence of small
imperfection. Classical examples include detection in nonwhite
Gaussian noise \cite{tree}
and image resolution beyond the diffraction
limit \cite{good}. Superposition of `macroscopic' quantum states is
supersensitive to loss \cite{cald}. This crucial sensitivity issue is one of
fundamental principle, not mere state of technology. It has thus far
received little attention in the field of quantum information.

As will be shown in sections V--VII, our qumode KCQ key generation
protocols are robust to channel parameter fluctuations. On the
other hand, the Lo-Chau protocol \cite{lo} is supersensitive at
high security level. This is because any amount of residue noise
in the system would be mistaken as E's action in the parity-check
hashing, and the protocol would be aborted according to its
prescription. The situation is particularly severe in view of our
discussion in III.C concerning E's ability to use her information.
The reverse reconciliation protocol in Ref. \cite{gros}, which
supposedly can operate in any loss, is supersensitive in high
loss. Let $\eta$ be the transmittance so that $\eta \ll 1$
corresponds to the high loss situation. In the presence of a small
additive noise of $\eta/2$ photons in the system, the protocol
becomes completely useless because the noise induced by the
attacker cannot be distinguished from excess noise. Apparently,
the modified Lo-Chau or Shor-Preskill type protocols can be
operated without such supersensitivity. Note that high security
level often decreases $P_{eff}$ and it is important to quantify
the tradeoff.

Even when the scheme is not supersensitive, the sensitivity level
has to quantified in a QKG scheme involving intrusion-level
detection for a complete protocol with quantifiable security, for
the following reason that has {\it not} been discussed in the
literature. The security proofs of such a scheme always has a
conclusion of the following form: If E has information $I_{E_i} >
\delta $ on $K^g$ from any attack $a_i$, then the probability that
$a_i$ would pass the test is $P_{E_i} < 1- \epsilon_{\delta}$ for
small $\delta$ and $\epsilon_{\delta}$. This may be put in form of
a conditional probability statement
\begin{equation}
\label{W} P(pass|I_{E_i} > \delta)< 1- \epsilon_\delta.
\end{equation}
However, the reverse conditional probability statement is required
for a real security proof:
\begin{equation}
\label{V} P(I_{E_i} > \delta|pass)< 1- \epsilon_\delta.
\end{equation}
It is easy to see that $P(I_{E_i} > \delta|pass)$ can be large
under (\ref{W}), say if E keeps attacking with $a_i$ that yields
significant $I_{E_i}$. For example, in standard BB84, E can learn
each bit with probability $\sim 0.85$ in an opaque attack. This
shows that the users must employ a `stopping strategy' in a
complete protocol, by adopting a stopping rule which stops the
whole process after a number of test results that lead to aborting
the protocol, and another rule to re-start the process. To
evaluate (\ref{V}), one would need $P(pass)$ in addition to
(\ref{W}) which would involve in turn
\begin{equation}
\label{U} P(pass| I_{E_i} < \delta).
\end{equation}
This probability (\ref{U}) depends on the quantitative sensitivity
level just discussed, corresponding to the case of no attack or
$I_{E_i}= 0$. In addition, E's optimal attack on the sequence of
key generation trials depends on the user's stopping and
re-starting strategy, and it appears one cannot bound the overall
$I(I_E > \delta| pass)$ before such a strategy is spelled out. It
should be clear that much remains to be done to obtain
quantitative results on the overall protocol security and
efficiency. In this connection, it may also be observed that in
all security proofs involving the use of an error correcting code,
it was not shown that an efficient (non-exponential) decoding
algorithm exists. Thus, on many levels it has {\it not} been
demonstrated that an efficient protocol exists with quantifiable
security levels even for standard BB84. On the other hand, these
problems do not arise in specific KCQ schemes.

After the protocol goes forward, there is clearly the question of
key-bit generation efficiency (or rate) $k^g_{eff}$ which may be defined
via the number of generated key bits subtracted by the number of key
bits used in the protocol, i.e.,
$K^g_{eff} \equiv (|K^g-|K|)/n $ when an $n$-bit data sequence was
used to obtain $K^g$ with a total key
$K= (K_s, K_c, K_m, K_v)$ that is not re-used.
 This should be distinguished from the final effective key
generation rate $k_r^g$, which includes all the operations in the
protocol to give the actual speed of key generation, a subject not
discussed in this paper.
When the modulation key $K_i$ with $|K_i|= |K_m|$ is not re-used
in the coded system of section III.4 in an $m$-block of $n$-bit
symbols,
\begin{equation}
\label{aah}
k^g_{eff} = R -|K_m|/n- I_E/n-|K_v|/mn,
\end{equation}
where $I_E$ is E's information rate that needs to be eliminated.
If the protocol is secure under known-plaintext attacks, the
$|K_m|$ term can be omitted. The $I_E$ term may be omitted if the
coding scheme automatically forces $I(X_n;Y^E_{n} K)$ to be
negligible.
\section{KCQ Coherent-State key generation with binary detection}\label{sec:kcqcoh}
In this section we describe the use of KCQ on qumodes,
quantum modes with infinite-dimensional Hilbert state spaces, for
key generation
via coherent states of intermediate or large energy. In most of the
current experimental developments \cite{gisi,gros} of QKG, coherent
states are employed in BB84 type protocols that are limited in energy
to $\sim 0.1$ photon, if only because of the photon-number splitting
attack that E can launch near the transmitter \cite{yue8,slut}.  With
KCQ, we will in this and the next section show that much larger energy
can be employed, line amplifiers and pre-amplifiers can be used, and
conventional optical technology on the sources, modulators, and
detectors can be utilized. Furthermore, direct
encryption
coherent-state KCQ in what is called the $\alpha \eta$ scheme has
already been experimentally observed
\cite{barb,corn}, which will integrate smoothly with the corresponding
key generation schemes that are currently under experimental
development.
\subsection{$\alpha \eta$ and its Extensions}
The usual description of a single coherent state already involves
an infinite dimensional space, referred to as a {\it qumode}.
Similar to the qubit case in Fig. 1, we may consider $M$ possible
coherent states $|\alpha_l \rangle$ in a single-mode realization,
\begin{equation}
\label{l}
\alpha_l = \alpha_0 (\cos \theta_l + i \sin \theta_l),
             \hspace{3mm}
             \theta_l= \frac{2\pi l }{M},
               \hspace{3mm}
                l \in \{1,...,M],
\end{equation}
where $\alpha_0^2$ is the energy (photon number) in the state, and
$\frac{2\pi l}{M}$ is the angle between two neighboring states. In
a two-mode realization, the states are products of two coherent
states
\begin{equation}
\label{m}
 |\alpha_0 \cos \theta_l \rangle_1 |\alpha_0 \sin \theta_l\rangle_2
 \hspace{2mm}, \hspace{2mm}
             \theta_i= \frac{2\pi l}{M} , \hspace{3mm}
               l \in \{1,...,M\},
\end{equation}
The qumodes may be those associated with polarization, time,
frequency, or any type of classical mode. Any two basis states
 form
a phase reversal keying (antipodal) signal set, which are nearly
orthogonal for $\alpha_0 \geq 3$. The optimal quantum phase
measurement \cite{pegg} yields a root-mean-square phase error
$\Delta \theta \sim 1/\alpha_0$. Thus, when $M\gg \alpha_0$,
the probability of error $P^E_b \sim 1/2$ when the basis is not known
which has been confirmed numerically [13], while
$P^B_b \sim \exp(-\alpha_0^2) \rightarrow 0$ when the basis is
known.

This scheme can be used for key generation as follows. An attacker
not knowing the key $K$ has to make a measurement to cover all
possible angles for different possible $K^{\prime}$ in her effort to
pin down the data $X$.
From such measurement result, she can then try to determine all the
possible $x$ corresponding to the different possible $k^{\prime}$. For
each running key $k^{\prime}$ from her trial key $k$ that selects
a particular basis for a particular bit, E has a classical binary decision
problem for two, not $m$, possible signal points.
 The more practical
heterodyne measurement, which may be forced on the attacker with
a signal set of varying amplitudes in addition to the  varying phase
of $\alpha \eta$, is
$6$ dB worse in energy than the optimal measurement \cite{hels}.
The optimal phase measurement, which has
no known physical realization, is worse than the optimal quantum
measurement of antipodal signals by $\sim 3$ dB  in signal energy
A detailed binary-decision numerical evaluation on this performance
is under way, but the $3$ dB estimate follows from the amplitude/phase
and conjugate
quadratures (heterodyne/homodyne) analogy, and
 is supported by known results \cite{hall}.
In any event, the precise number is important in an actual design
of real system and in bringing out the intrinsic limitation of the
system, but is not as important for illustrating the possibility
and basic principle involved. In this case, the principle is that
there is a substantial difference in performance due to a {\it
quantum effect} that has no classical analog, viz, different
incompatible quantum measurements versus a single complete
measurement in the classical case.

More precisely, for discrimination of two equally likely coherent states
$\{|\alpha_0\rangle, |-\alpha_0\rangle \}$, the optimum quantum receiver yields
an error rate $\bar{P}_b$ that may be
compared to the heterodyne result $P^{het}_b$
and the phase measurement result $P^{ph}_b$, with $S= \alpha^2_0$,
\begin{equation}
\label{C}
\bar{P}_b= \frac{1}{4} e^{-4S}, \mbox{     }
P^{het}_b \sim \frac{1}{2} e^{-S}, \mbox{     }
P^{ph}_b \sim  \frac{1}{2} e^{-2S}
\end{equation}
Here, $S$ measures the average number of photons received in the
detector and (\ref{C}) applies in the so-called quantum-limited
detection regime--- unity detector quantum efficiency, infinite
detector bandwith, all device noise suppressed. Under (\ref{C})
and dropping the factors in front of the exponentials for a
numerical estimate of the bit-error rate (BER), which is required
to be $\leq 10^{-9} $ per use in a typical communication
application, we have, for a mesoscopic level $S \sim 10, \bar{P}_b
\sim 10^{-12}, P^{het}_b \sim 10^{-3}, P^{ph}_b \sim 10^{-6}$. If
the data arrives at a rate of $1$ Gbps, the user B is likely to
have $10^9$ error-free bits in $1$ sec, while E would have $\sim
10^3 $ errors among her $10^9$ bits with the optimum phase
measurement. By the usual privacy distillation approach
\cite{ben3}, the users can generate $\sim 10^{3}$ secure key bits
by eliminating E's information. Thus, in principle, $\alpha \eta$
in its original form is capable of secure key generation against
individual attacks that employs the optimal phase measurement on
each qumode. Similar to {\it all} cases of specific QKG schemes to
date, that there is no full security proof against even constant
individual attacks in contrast to the claim of unconditional
 security in existence proofs,
 the above analysis does not prove there
is no other individual or collective measurement, particularly
adaptive ones utilizing the seed key information, that would yield
a substantially better BER for E than the optimal phase
measurement. Intuitively, we feel that is quite unlikely, but new
techniques in classical and quantum detection theory are being
developed to give precise quantitative treatment on such problems.
Note that whatever the final result may turn out to be, it only
affects the quantitative advantage level but not the possibility
of advantage creation. In this connection, we may mention that
$\alpha \eta$ in its original form was proposed for key generation
in Ref. \cite{bar2}, with no consideration of
information-theoretic key security against meaningful attacks. In
the way $\alpha \eta$ was run, actually no
 fresh key can be generated
because $P^{het}_b$ in (\ref{C}) is very small.

More serious limitations on the use of $\alpha \eta$ for key
generation, I believe, arise from the US Advantage Creation
Principle when the above scheme is to be utilized in practice. In
the first place, device thermal noise is significant at high data
rate and small signals, thus optical pre-amplifiers need to be
used. For the usual erbium amplifier this would already take out
the advantage over E. On this issue, it may be pointed out that
the optimum binary quantum receiver has not been implemented so
far in $\alpha \eta$, but the near-optimum Kennedy receiver
\cite{kenn}, with $P^{\prime}_b= (1/2) e^{-4S}$ is currently under
development. (The factor $1/2$ difference between $\bar{P}_b$ and
$P^{\prime}_b$ can be recovered in a Dolinar receiver described in
Ref. \cite{doli},
 which is the first systematic
investigation of optical receiver performance improvement via
 feedback.)
On the other hand, the photon number amplifier (PNA)
\cite{yue8,yu10} could lead to an ideal Kennedy receiver in
principle although PNA is far from practical at present. Secondly,
in the presence of a line loss $\eta$  from A to B, one would need
to compare $P^B_b \sim e^{-4 \eta S}$ to $P^E_b \sim e^{-2S}$
according to the Advantage Creation Principle when E attacks near
A. Even if one uses the advantage creation technique of accounting
for E's energy splitting described in III.G, and the postdetection
selection technique of Ref. \cite{yuen}, in conjunction it appears
difficult to create an advantage over E that would allow key
generation over truly long-distance telecomm fibers. Thus, a more
powerful approach via $m$-ary detection is developed in the
following. Before we turn to this advantage creation technique, it
is useful to introduce a number of other techniques that may
improve the security and efficiency of KCQ schemes, and to
demonstrate a general limitation on the binary detection approach
to coherent-state KCQ for key generation.

It is important to note that $\alpha \eta$ represents a new type
of cipher even when it is operated in a completely classical
setting, and even in the absence of any channel noise. This is
because deliberate randomization may be introduced by A in many
ways. Consider the situation where the circle of Fig.1 represents
a classical two-dimensional signal space, say corresponding to the
two quadratures of a single frequency. In the absence of any
noise, the circle would represent the different possible
phase-shifted signals of a given energy. Thus, the cryptosystem
can be run in the same way as $\alpha \eta$ with or without
classical noise, in hardware and even in software. Indeed, in the
direct encryption experiments on $\alpha \eta$ reported thus far
\cite{corn,cove}, the performance obtained via the coherent-state
quantum noise can also be achieved by high-speed deliberate
partial randomization of the signals by A corresponding to the
coherent-state noise effect. When $\alpha \eta$ is used for key
generation, one may employ the technique of {\it deliberate signal
randomization} (DSR)
 for which each signal state at the output of the
encryption box of Fig. 1 is further randomized so that it is
 uniformly distributed on the
semi-circle centered at the state chosen by the running key
$K^{\prime}$. Similar to the qubit case, one may readily
 show the intuitively obvious fact that the key
$K$ is completely hidden form E who does not know $K$ and $X$,
 even if she possesses one copy of $\rho^k_x$ corresponding to
an arbitrarily long data sequence $x$. The general interwined case
described in section III will be treated in Part II, as it
involves security against known-plaintext attack on direct
encryption. This is true classically also as just discussed. In
the presence of quantum or classical noise, one needs to use a
proper randomization if all the error control is built in the
antipodal signal set only, as above. When a CECC is used on top of
the antipodal signals, deliberation error randomization can be
introduced to improve security/efficiency as described in section
III.G.

In addition to providing complete protection against ciphertext-only
attack on the key, DSR also improves the efficiency of key generation.
If the state is rotated by an angle $\theta$ away from the one set by
$K^{\prime}$ and is unknown to the detector which knows $K^{\prime}$,
the optimum phase measurement BER as a function of $\theta$ is yet
to be evaluated, but the corresponding Kennedy and heterodyne
receiver performance are
\begin{equation}
\label{D}
\bar{P}_b(\theta) \sim  e^{\frac{-2S\cos^2 \theta}{(1-\cos)^2} },
\mbox{         }
P_b^{het}(\theta) \sim \frac{1}{2} e^{-S\cos^2 \theta}.
\end{equation}
The $\bar{P}_b(\theta)$ in (\ref{D}) is the
Chernov bound \cite{woze}
for the Kennedy receiver when $\cos \theta \ll 1$,
while $P_b^{het}(\theta)$ is the usual
upper bound on Gaussian errors \cite{wyne}.
It is expected that in a more exact evaluation of
 $\bar{P}_b(\theta)$, the energy advantage is closer to the
 original $6$ dB than the $3$ dB one of (\ref{D}), and similarly for
 $P_b^{ph}(\theta)$. Thus, even for large signal energy, A can control
 the BER to B and E and causes more errors to E through B's
 advantage. In practice, some CECC should be used for reliable
 system operation, and channel code key $K_c$, chaining, and other
 techniques could be used to enhance the advantage already created
 for efficiency improvement.

 With DSR on $\alpha \eta$, one may obtain secure key generation
 against constant individual attacks as follows. Let the angle $\theta$
 be randomized so that it appears uniform over the whole circle with
 respect to E's optimum (phase) constant qumode measurement. In
 this way, the key $K$ is completely hidden even in a known-plaintext
 attack, in a way exactly similar to the classical noiseless case where
 a semicircle is sufficient to protect against ciphertext-only attacks.
 If the signal strength $S$ is not large enough, this would also introduce
 error to B. However, a CECC can be designed to correct only up to
 the BER B then suffers, which is smaller than that of E due to B's
 error
 performance advantage. By using Theorem 1, one may generate
 fresh keys, with no cost for each $n$-bit symbol due to security
 against known-plaintext attacks. We summarize

\vspace{5mm}
{\it Theorem 5}:

With proper use of DSR just described, a coded $\alpha \eta$ scheme
leads to unconditionally secure net key generation against constant
individual attacks with security level given by (\ref{B}).

\vspace{5mm}

Again, with the development of proper bounding techniques, we
believe the restriction to constant individual attacks in Theorem 5 can
be simply removed. The detailed quantitative dependence of the security
level as a function of $S$ and other system parameters will be given
for both individual and joint attacks in the future.
\subsection{Binary Detection KCQ Key Generation}
For binary coherent-state signals, the optimal quantum receiver
performance cannot be better than that of heterodyne by $6$ dB in
energy or error exponent. This is a known fact among all the usual
binary coherent state systems, but there is no general proof in the
literature. A proof can be supplied, which is not difficult, but is
omitted here for brevity. Also, it may be proved that antipodal signals
lead to optimal BER under energy constraint on coherent states.
Furthermore, it is not possible to increase the error advantage
by utilizing bandwidth, or more generally any multimode system, for the
following reason.

Consider an optical quantum field of arbitrary bandwidth $E(x,t)$ where
$x$ is the transverse spatial dimension. On her copy of the
field, whether it is the one she splitted off by tapping or the
hypothetical one we grant her for bounding her information, she can
always in principle make a heterodyne measurement to obtain the
classical readout $\varepsilon(x,t)$, which is described by \cite{yu11},
\begin{equation}
\label{E}
\varepsilon(x,t) = \varepsilon_s (x,t) + n(x,t),
\end{equation}
where $\varepsilon_s (x,t)$ is the amplitude of the coherent-state
signal and $n(x,t)$ is an additive Gaussian noise in $t$ with spectral
density $hf$ at frequency $f$.
All quantum fluctuation in every space-time mode has already
 been included.
In a binary detection system involving two classical signals in additive
white Gaussian noise (AWGN), one can always extract one signal
dimension (one quadrature out of one mode) that contains all the
information for optimal discrimination \cite{woze}.
 The whiteness approximation
that all $f\sim f_0$, a single frequency, is very good for all
practical optical signals. Thus, even though (\ref{E}) contains
many quantum noise photons-- one from every mode-- in the optimal
receiver only one such mode is to be extracted by appropriate
spatial-temporal filtering. As a consequence, we are back to the
single-mode
 situation where there is just one noise photon from heterodyne.
Indeed, heterodyne is a `universal' measurement whose result
captures all aspects of the field mode: amplitude, phase, quadratures,
etc., that allows E to try all possible $K$ to identify all possible data.
One may understand this result from the important fact that
a multimode coherent-state excited field from vacuum is equivalent to
a single-mode coherent-state excited field.

The result just quoted is not true for {\it nonclassical light},
i.e., optical fields in quantum states that is not classical, not
a coherent state or a classically random superposition of coherent
states. Clearly, there can be huge improvement between the optimal
and heterodyne detection of a nonclassical state. For number
states, ideal photon counting yields $P_b=0$ for on-off signals.
For squeezed states, homodyne detection along the maximum
squeezing direction sees the minimum noise as compared to one that
may see a large noise without knowledge of that direction. Thus,
by using $K$ to determine such directions, the users would obtain
huge error advantage over E even in a binary detection system. One
can similarly use number states and other orthogonal states in
conjunction with coherent states to create other binary systems
that give arbitrarily small BER for B but large ones for E.

We would not go into the details for such development because
intermediate or large-energy nonclassical states do not have much
practical significance as data source in long-distance
communication \cite{yue3}. This is because the inevitable system
imperfection, especially linear loss, would quickly transform such
nonclassical states into classical ones. As a consequence, the
initial energy or error advantage disappears quickly over a lossy
communication line. For realistic application of mesoscopic or
macroscopic energy signals, we may want to limit ourselves to
coherent states.
\section{KCQ coherent-state key generation with $m$-ary detection}\label{sec:kcqmary}
The above limitation on the binary detection advantage of an optimal
quantum receiver versus heterodyne can be overcome in $m$-ary
detection. The use of $m$-ary systems, in fact, is one form of
coding. As will be seen in the following, it indeed corresponds to
driving the system at a rate between B's and E's mutual information
with respect to A as in (\ref{aad}). Amazingly, for the particular
CPPM system we now turn, such a rate choice by A automatically
makes $I_E$ go to zero with a flat error profile, with also full
information-theoretic security against known plaintext attack on the
key. This is proved against the universal heterodyne attack, and is likely
to be true against all possible attacks. Thus, not only the data enjoy
unconditional security at the near perfect level, the key has security
that has never even been suggested possible before
 in either standard or quantum cryptography.
\subsection{CPPM--- Coherent Pulse Position Modulation}
An $m$-ary coherent-state pulse position modulation system has the
following signal set for $m$ possible messages,
\begin{equation}
\label{H}
|\phi_i\rangle = |0\rangle_1 \cdot\cdot\cdot
                   |\alpha_0\rangle_i \cdot\cdot\cdot
                   |0\rangle_m,
                   \mbox{          }
                   i \in \{1,...,m \}.
\end{equation}
In (\ref{H}), each $|\phi_i\rangle$ is in $m$ qumodes all of which
are in the vacuum state except the $i$th mode, which is in a
coherent state $|\alpha_0\rangle _i$. The corresponding classical
signals are orthogonal pulse position modulated if each mode is
from a different time segment, but generally the modes can be of
any type. For brevity, we retain the term `pulse position' even
through `general mode position' is more appropriate.

The photon counting as well as heterodyne error performance of
(\ref{H}) are well known \cite{gagl}. The block error rate from direct
detection is exponential optimum for large $m$.
\begin{equation}
\label{I}
P^{dir}_e=(1-\frac{1}{m}) e^{-S}, \mbox{      }
\bar{P}_e \rightarrow e^{-S}.
\end{equation}
The optimum block error rate $\bar{P}_e$ for (\ref{H}) is known
exactly \cite{yue2}, and given in (\ref{I}) asymptotically.
 In contrast, for large $m$ the heterodyne block error rate
$P^{het}_e$ approaches $1$ exponentially in $n= \log_2 m$, which is
a general consequence of the Strong Converse to the Channel
Coding Theorem as discussed in section III.D.
For the present Gausssian channel case for heterodyne receivers,
explicit lower bound on the block error rate $P^{het}_e$, conditioned
on any transmitted $i$, can be obtained in the form
(p382 of \cite{gag2})
\begin{equation}
\label{J}
P^{het}_e > (1-[\Phi(y)]^n) \Phi(y-\sqrt{2S}),
\end{equation}
where $\Phi$ is the normalized Gaussian distribution. By choosing
$y > \sqrt{2n}$, (\ref{J}) yields explicitly
$P^{het}_e \rightarrow 1$ exponentially in $n$ for any given $S$.
It is a main characteristic of classical orthogonal or simplex signals
in AWGN that whenever an error is made, it is equally likely to be
decoded by the optimal receiver to any of the $m-1$ other messages.
Thus, under the condition $P^{het}_e \rightarrow 1$, the error
profile is uniform, viz, $p_i = 1/m$ or the BER $P_b= 1/2$ with
independent errors.

The KCQ qumode key generation scheme CPPM works as follows.
Consider $m= 2^n$ possible $n$-bit sequences, and possible
coherent-states
\begin{equation}
\label{K}
|\psi_i\rangle = \otimes^m_{j=1} |\alpha_{ij}\rangle^{\prime}_j,
\hspace{5mm}  i,j \in \{1,...,m\}
\end{equation}
in correspondence with $\{|\phi_i\rangle \}$ of (\ref{H}).
For simplicity, one may set $\sum_j |\alpha_{ij}|^2= |\alpha_0|^2= S$
for every $i$. Let $f_k$ be a one-to-one map between (\ref{H}) and
(\ref{K}) indexed by a key $K$. As an example of physical realization,
the connection between (\ref{H}) and (\ref{K})
could be through a set of $N$ beam-splitters with
transmission coefficients $\sqrt{\eta_l}$ for complex numbers
$\eta_l$, $l \in \{ 1,..., N \}$, determined by $k$. Such a physical
realization combines the $\alpha_{ij}$ of (\ref{K}) coherently through
the $\eta_l$'s, and is represented by a unitary transformation
between the two $m$-tensor product state spaces
$\otimes^m_{i=1} H_i$ and $\otimes^m_{i=1} H^{\prime}_i$ for
the input and the output.
\footnote{Note that this description is neither the Schrodinger nor the
Heisenberg picture, but is more convenient in problems of quantum
system analysis.}
The states $|\psi_i \rangle$ of (\ref{K}) are used to modulate the
data $i$ by A, and B demodulates by first applying $f_k$ to transform
it to $|\phi_i \rangle $ of (\ref{H}) and then use direct detection
on each of the $m$ modes $H_i$.

Without knowing $f_k$ or $\eta_l$ so that there are both amplitude
and phase uncertainties for each $l$, it is expected that an attacker
can do very little better than heterodyne on all the $H^{\prime}_i$
 modes,
which is equivalent to heterodyne on all the $H_i$ modes, and then
apply the different $f_k$'s on the classical measurement result
(\ref{E}). As presented above, by making $m$ large one can then
make not only $\bar{p}_E = 2^{-l}$ for any $l$ but E's error
profile is in fact nearly uniform, with $p_i = (1-2^{-l})/(m-1) $
for $i \geq 2$, thus no need for further privacy distillation. As
a consequence, the system is not only completely secure against
ciphertext-only attack on the key but also fully secure against
known-plaintext attacks. This is because given an input-output
pair $(X_n, Y^E_n)$, the heterodyne output $Y^E_n$ has no relation
to $X_n$ for any $k$ from E's uniform error profile. The exact
quantitative behavior may be bounded via (\ref{J}). We summarize:

\vspace{5mm}
{\it Theorem 6}

Against E's universal heterodyne attack, the $m$-ary CPPM KCQ
protocol is unconditionally secure with asymptotic key generation
rate $n= \log_2 m$ per use and $\bar{p}_E$, $I_E$ going to zero
exponentially in $n$.
\vspace{5mm}

The only easy way to remove the restriction to heterodyning for E is to
note that the optimum quantum receiver for discrimination among the
states (\ref{H}) is unique \cite{ken2}. Thus, there is a gap
 between it and the
receiver performance that does not know $K$ at the time of quantum
measurement, which translates into a mutual information statement
(\ref{aad}) that can be used to show the existence of codes upon
further coding, as described in III.E, that yields security in the
sense of Theorem 1. Although this does not seem to give useful
practical protocol for actual implementation and does not
guarantee key bit generation, it is of interest in principle to
record the following.

\vspace{5mm}
{\it Theorem 7}:

Against any attack by Eve, the CPPM scheme may be further coded
to provide unconditional security with levels given by (\ref{B}).
\subsection{Further Outlook}
The direct detection or optimal detection performance (\ref{I})
is affected by the presence of device noise so that there is
 no more vacuum
state in (\ref{H}). However, ordinary pre-amplifier
 could be used that suppresses all the device noise with
a resulting performance degradation that amounts to a less factor
$1/4 \leq \eta < 1$ in $m$-ary PPM. Furthermore, in principle a
photon-number amplifier mentioned in VI.A can be used as a
noiseless pre-amplifier. Quantitative evaluations of the resulting
performance are, however, yet to be carried out. One major
advantage of coherent-state KCQ scheme is that they can be used
through a limited number of amplifiers and switching nodes in a
properly designed system with appropriate amplifiers. In general,
quantum amplifiers degrade the user's error performance due to the
fundamental quantum noise they introduce \cite{yue8}.
 In a properly
designed chain with appropriately chosen amplifier gains and lengths
of lossy line segments, one can obtain a linear \cite{yue6}
instead of
an exponential degradation in the signal-to-noise ratio (SNR) as
a function of total line length. There is no need to decrypt and
re-encrypt at the input of an amplifier as in a repeater, as long as
the degradation introduced by the amplifier still leaves B with
performance advantage over E. Recall the overall general Advantage
Creation Principle for key generation that
B must have performance advantage on the decoded information-bit
sequence after all system imperfections including loss and noise are
taken into account,as compared to E's decoded information-bit
sequence for no loss and no imperfection other than unavoidable
ones.
In the case of CPPM, B's performance would be scaled by the total
transmittance $\eta$ so
that $S$ is replaced by $\eta S$ in
(\ref{I}). In principle, it is still a better performance compared to
(\ref{J}) with $\eta =1$ for large enough $m$. Thus,
CPPM can be secure for arbitrarily long-distance fiber communication.

It may be mentioned that the possible use of amplifier in a quantum
cryptosystem has been introduced previously for weak coherent
states and heterodyne/homodyne detection
\cite{yue5} that traces back
to Ref. \cite{yue7} that describes both coherent-state and
squeezed-state cryptosystems.
In particular, the usual-state scheme in
Ref. \cite{yue5}
employs conjugate-variable measurement detection of intrusion level
 similar
to the schemes of Ref. \cite{gros}, while also allowing a limited use of
amplifiers as described. However, all such schemes are
inefficient because weak or small energy signals have to be used
to avoid good performance in determining the actual signal state
via optimal quantum detection by E, and
 via attacks similar to the
USD attack on coherent-state realization of BB84
type systems \cite{yue8,slut}.

The CPPM scheme is also ideal for direct data encryption because it
automatically produces a near uniform error profile on E corresponding
to near-perfect bit-by-bit security. Indeed, from the constant
inner product $\langle \phi_i | \phi_j \rangle$
for every $i \neq j$ which is the quantum analog of the classical
orthogonal or simplex signal behavior that is responsible for their near
uniform error profile in AWGN,
 it would be possible to prove that such a property
persists under E's optimal attack. It would then appear that all problems
are solved in principle as arbitrarily large error exponent advantage
can be obtained between (\ref{I}) and (\ref{J}) by making $m$
large.

Unfortunately, as in a classical orthogonal signaling
 scheme, large $m$ in CPPM means
exponential growth of bandwidth, not to mention growth in physical
complexity. Indeed, (\ref{I}) itself is an infinite-bandwidth
result for large $m$. One the other hand, it is known \cite{yue3}
that if the signal-to-quantum noise per unit bandwidth is small,
coherent-state direct detection systems do have larger capacity
than heterodyne ones. Thus, it may be expected that properly
designed error correcting codes, usually employed for bandlimited
systems for such reasons, could be developed to retain much of the
CPPM
 advantage for a large given bandwidth.
\section{KCQ and direct encryption}\label{sec:kcqdirect}
For direct encryption, one needs to consider ciphertext-only
attack on the key, on the data, and known-plaintext attack on the
key. In conventional cryptography one has the Shannon bound,
$H(X|Y) \leq H(K)$, on the conditional entropy of the data $X$
given the ciphertext $Y$ via the key entropy.
 In the quantum case or in the presence
of irreducible classical noise to E, the corresponding bound
\begin{equation}
\label{n}
H(X|Y_E) \leq H(K)
\end{equation}
is no longer valid where $Y_E$ is the classical ciphertext available to E.
In the quantum case, $Y_E$ is obtained via  a quantum measurement.
 If (\ref{n}) is valid as is the case in conventional cryptography,
 it does not mean that E knows all the bits in $X$ except for
$|K|$ of them. That would be disastrous as it often happens that
$|K| \leq 10^3$ while $|X| > 10^9$. The operational meaning of
 (\ref{n}) has never been analyzed in conventional
cryptography, to my knowledge. It is usually not considered a
problem because it is presumed that E would get many information
bits in $X$ wrong knowing only $Y$ and not $K$. However, a more
detailed analysis is needed for a security proof with respect to
whatever chosen criteria, as we have done for key generation in
section III.C. But that has never been provided in conventional
cryptography other than the trivial one-time pad case. When
(\ref{n}) is violated, Lemma 1 implies $I(X;Y_E K) < H(X)$, a
condition that allows key generation via (\ref{aab}) as in Theorem
1. For direct encryption, such violation has the {\it important}
implication that very high level of data-bit security may be
obtained {\it without} using the inefficient one-time pad. Indeed,
we have seen how this may occur in CPPM treated in section VI.
Note that Theorems 1 and 2 can also be used to describe the data
quality in direct encryption.

It is easy to protect $K$ from ciphertext-only joint attacks in
$\alpha \eta$ with the use of, e.g., DSR, discussed in section
V.A. The technique can be extended to cover known-plaintext
attacks in two different ways, to be presented in Part II. Without
DSR, the bound (\ref{n}) obtains with $H(K)/\log M$ on the
right-hand side. The usual security problem is known-plaintext
attack, in which E tries to determine $K$ from data-output
sequence pairs with statistical correlation information on the
data (of varying degree). Security against known-plaintext attacks
 is always at best computational complexity-based
against exponential search in conventional cryptography. For noisy
system, we suggest that it is possible to have
information-theoretic security, i.e.,
\begin{equation}
\label{M}
H(K|X,Y_E) > 0, \mbox{     } H(K|X, Y_B) = 0,
\end{equation}
which has never been suggested before and is clearly impossible in
conventional cryptography where $Y_E = Y_B$. Full security would
correspond to $H(K|X,Y_E) = H(K) $, which again can be closely
approximated in CPPM systems, at least for the universal
heterodyne attack.

Even when $H(K|X,Y_E)=0$ as is the case in conventional
cryptography for sufficiently long $X$, the system may be secure
in the sense of high search complexity. In particular, $\alpha
\eta$ in the original form without DSR provides an additional
search problem to E, as compared to just the encryption box, that
is exponential in $|K|/\log_2 M$, at least for brute-force search.
For increasing the search complexity, one may make sure the input
data can never be perfectly known in several ways. One is to use
polarity or padding bits as described in Section II.
 Another is to generate the polarity bits
through a running key obtained from another encryption mechanism
with the same $K$ or another different key as the seed key. Note
that the proper use of DSR would introduce inevitable
coherent-state quantum noise for E, which may even lead to
information-theoretic security already if the energy in the
coherent state is not too large. No proof of any cryptosystem has
ever been given in conventional cryptography, to my knowledge,
that establishes rigorous exponential lower bound on the search
complexity.
 And we have not (yet) succeeded in
proving that $\alpha\eta$ necessitates  an exponential search either -
it is just an added search burden as compared to just the encryption
box and appears exponential. In general, such multi-variable
correlated classical statistical problem has the full mathematical
complexity of many-body problems and quantum field theory in
physics. Useful lower bound is also notoriously difficult to obtain in
computation problems.
 Perhaps these explain why no rigorous security proof
is available on such complexity-based security.

On the other hand, exponential search complexity should be good
enough for any application. We have mentioned in section II that
Grover's search only reduces the exponent by a factor of two,
which is easily compensated by increasing the key size by a factor
of two in many standard schemes as well as in our KCQ schemes,
either qubit or coherent-state.

As noted previously in this paper, data security against attacks
on the key with statistical knowledge on data that are not
completely random is required for a complete proof of key
generation security with KCQ. An extensive general theoretical
development of direct encryption security analysis will be
provided in Part II, where conditional probabilities will be used
in addition to entropies for more precise quantitative estimate of
specific coding/detection scheme performance. The behaviors of
$\alpha \eta$, CPPM, and their refinement and extension will be
developed for concrete cryptosystem design.
\section{Comparison among QKG schemes}\label{sec:qkgcomp}
We present below a brief qualitative comparison between QKG
schemes of the BB84 type, of the qubit KCQ type, and the
coherent-state KCQ type. Detailed quantitative comparisons will be
given after rigorous evaluation of the quantitative characteristics of
these schemes for finite $n$.

Theoretically, BB84 type protocols suffer from the following classes
of problems as a matter of fundamental principle.

\vspace{3mm} (i) It is hard to bound Eve's error rate on the key
generated due to the difficulties of intrusion-level estimation
under joint attacks with side information on the error correcting
and privacy distillation codes.

(ii) It is difficult to produce a complete protocol that can be
practically implemented with quantifiable security and efficiency,
due to the decoding problem and the stopping-rule problem.

(iii) It is hard to include the various system imperfections in an
unconditional security proof, and to build protocols robust with
respect to fluctuations in the magnitude of these imperfections.

(iv) The necessary use of weak signals and the difficulty of
repeating the signal without decryption imply low throughput even
with just moderate loss.

(v) The intrinsic small quantum effect of a single photon
necessitates an accurate sensitivity analysis with respect to the
system imperfection and environmental perturbations, that would
result in a low interference tolerance threshold in commercial
applications. \vspace{3mm}

Corresponding to these problems are related practical one including

\vspace{3mm} (i$^{\prime}$) It is difficult or impossible to
rigorously ascertain the quantitative security level of the
generated key.

(ii$^{\prime}$) The throughput or key-generation rate would be low,
especially in the presence of substantial loss.

(iii$^{\prime}$) The cryptosystem is sensitive to interference, and
needs to be controlled and checked with a high precision difficult to achieve
practically.

(iv$^{\prime}$) High-precision components corresponding to
a fundamentally new technology are required, including the source,
transmission line or repeater, and detector.

\vspace{3mm}

With the exception of (i) and (ii), which are further discussed in
Appendix A, all these problems are evident and well known in BB84
although there are disagreements on how readily they can be
overcome. Nevertheless, in the foreseeable future it seems clear
that BB84 type schemes cannot be made to operate in a commercial
type environment with any reasonable level of security and
efficiency for even moderately long distance. The weak
coherent-state schemes of Ref. \cite{gros} also suffer from all
these problems except (iv$^\prime$),
 and that of Ref.
\cite{yue5}
is only slightly better.

With the use of qubit KCQ type schemes, the theoretical problems
in  (i) and (ii) can be largely overcome, but not the ones in (iii)
and (iv) except perhaps with very low key-bit generation efficiency.
Except for (i$^{\prime}$), the practical difficulties
(ii$^{\prime}$)-(iv$^{\prime}$) also remain. Again, it may be possible to
alleviate these problems with strong error correction that implies
a low $k^g_{eff}$.

With the use of the qumode KCQ schemes of intermediate to large
energy, all the fundamental difficulties (i)-(iv)
can be substantially reduced. Furthermore, each of
 the practical difficulties
(i$^{\prime}$)-(iv$^{\prime}$) either disappears or is substantially
alleviated. The exception is loss in long-distance fiber communication.
In principle, a wideband coherent  CPPM system presented in section
VI could solve all problems. For practical application,
 new approaches are needed
to deal with bandwidth limitation and coherence requirements.
It is still a major problem to create enough advantage
for unconditional information-theoretic security.
\section{Concluding remarks}\label{sec:concl}

A new principle of quantum cryptography has been presented on the
basis of optimal versus nonoptimal quantum detection when a seed key
is known or not known. This possibility of yielding better performance
for the users over an attacker is a quantum effect with no classical
analog. In classical physics, a complete observation of the physical
signal state can be made with or without the key. It would be
misleading to phrase the basis of this possibility
 as no-cloning, which is trivially covered
\cite{yue9,dari}
by quantum detection theory that provides detailed quantitative limits
on quantum state discrimination from the laws of quantum physics.
A detailed development of the appropriate novel quantum
detection theory wil be given in the future for a complete quantitative
assessment of cryptosystem efficiency and security.
This will be done especially in terms of Eve's optimal probability of
guessing the generated key correctly, which is a more appropriate
criterion than her mutual information.

A powerful new KCQ protocol CPPM that utilizes $m$-ary instead of
binary detection has been presented that could, in principle, lead
to secure key generation and data encryption over long-distance
telecomm fibers. However, the problem of obtaining such a protocol
under bandwidth and practical constraints remains both a
theoretical and experimental challenge.
\acknowledgments I would
like to thank E. Corndorf, G.M. D'Ariano, M. Hamada, O. Hirota,
W.-Y. Hwang, P. Kumar, H.-K. Lo, N. L\"utkenhaus, R. Nair, M.
Ozawa,
 and M. Raginsky for useful
discussions on cryptography as well as many topics in this paper.
This work was supported by the Defense
Advanced Research Project Agency under grant F30602-01-2-0528.

\appendix
\section{Problems of QKG unconditional security proofs}\label{app:qkgprobs}
In the QKG literature, there are different types of security
proofs that purport to show the existence of unconditionally
secure protocols against any attack allowed by the laws of
physics, while new proofs are continually emerging that we have
not scrutinized. They are the proofs of Shor-Preskill \cite{shor},
Lo-Chau \cite{lo}, Mayers \cite{maye}, Biham etc. \cite{biha} and
recently a more complete approach by Hamada \cite{hama}. I believe
that all of them have serious gaps that are difficult to close.
The situation is rather confusing because the strategies of these
proofs are different, and problems in one type may not arise in
another. In the following, three major problems in these proofs
are briefly described, at least one of them applies to any of the
above proofs. In addition, three other problems in a complete
protocol are briefly indicated, none of which has been addressed
in the literature, to my knowledge. \footnote{ While only I am
responsible for the assertions in this Appendix, they are made
after extensive discussion in our group that include also G.M.
D'Ariano, W.-Y. Hwang, R. Nair, and M. Raginsky who made important
contributions and clarifications that make the writing of this
Appendix possible. I also benefited from exchanges with M. Hamada,
H.-K. Lo, and N. L\"utkenhaus. I hope this Appendix would
stimulate serious exchanges, and that it would be replaced by
separate papers in the future. }.

The three major problems in these proofs are

\vspace{3mm} (i) It is hard to rigorously and accurately estimate
E's disturbance on the qubits under a joint attack.

(ii) It is difficult to show there exists a universal error correcting
code that would produce the desired security level for all attacks
that pass the intrusion-level test in the protocol with significant
probability.

(iii) The side information that E has from the public knowledge has
not been properly taken into account in the estimate of her
information on the key generated.

\vspace{3mm}

From an information-theoretic viewpoint, a joint attack from E
creates in general a quantum
channel with entanglement on the user's qubits.
It seems impossible to obtain a good estimate
of E's disturbance from merely one copy of the general channel.
This difficulty, which may be called the {\it inference problem},
 also affects the existence and the choice of a code
that would perform satisfactorily under all possible attacks that
do not lead to the protocol being aborted, as expressed in (ii)
above. Indeed, even for constant individual attacks this {\it
coding problem} has only recently been rigorously dealt with in
Ref. \cite{hama}.

When it applies, the inference problem is serious and does not
seem to have a solution even in the asymptotic limit $n
\rightarrow \infty$. It seems to be handled in Ref. \cite{shor} by
the argument of quantum to classical reduction adopted from Ref.
\cite{lo}. However, the local measurement actually performed by
the users is {\it not} equivalent to the nonlocal degenerate Bell
measurement needed for the reduction to go through, with respect
to the determination of the state after the measurement that is
 needed in the next step of the proof.
They are only equivalent, in both the cases of Ref. \cite{shor}
and Ref. \cite{lo}, with respect to the probabilities that govern
the use of the measurement results. Perhaps the equivalence claim
arose from interpreting the description of a measurement by the
same $X\otimes X$ differently in two different contexts
\cite{got2}. When $X\otimes X$ has a degenerate spectrum, the
measurement as specified by a POM is not uniquely represented by
the symbol $X\otimes X$. Furthermore, the inference of the test
qubit results to the information qubits left cannot be justified
by the quantum de Finetti Theorem \cite{lo3,cave} because quantum
entanglement leads to violation of the exchangeability premise of
the theorem, and quantum entanglement is precisely what a joint
attack can yield that an individual attack cannot. However, the
inference problem does not arise in proofs where B makes
measurements on all received qubits before proceeding. The problem
of such proofs is how one may bound Eve's information under her
optimal attack.

Note that a security proof needs to answer this question for the
user: given that a key $K^g$ is obtained by following the protocol
on $n$ qubits, what one can rigorously say about the error profile
or information that E has on $K^g$ as optimized over all her
possible attacks. This question is especially serious for the
realistic case when the statistical fluctuation due to a finite
$n$ needs to be under control. It appears that new techniques need
to be developed to handle such problem in this type of protocols
with intrusion-level detection. Even asymptotically, the {\it
coding problem} remains on what scheme one should employ that
guarantees a bound on Eve's information when she optimizes her
probe/interation {\it in anticipation} that she would receive side
information later before she makes her measurement. This problem
is coupled with the following side information problem, although
the latter constitutes a problem al by itself.

In terms of our notation, the {\it side information problem} can
be simply stated as follows. Let $S$ be E's side information
before she made her final measurement and estimate of the
generated key $K^g$ from an observation on her ancilla. Then Eve's
mutual information on $K^g$ is given by $I(K^g; Y_E S)$. From
(\ref{Z}), this is equal to $I(K^g; Y_E| S)+ I(K^g; S)$. Most
treatments just bound $I(K^g; Y_E)$. In \cite{hama}, the (smaller)
$I(K^g; Y_E| S)$ is bounded but $I(K^g;S)$ is ignored. However,
$I(K^g;S)$ may grow with the number of qubits $n$ in $x$ and has
to be subtracted from $K^g$ to show that a net positive key
generation rate is in fact obtained. In this regard, we may recall
that the fundamental superiority of quantum over standard
cryptography is based almost exclusively on the availability of
rigorous proofs rather than mere plausible assumptions,
qualitative arguments and numerical simulations.

The three problems that need to be dealt with in a complete
protocol that be be practically implemented with quantifiable
security are \vspace{3mm}

(1) The Efficient Code Problem:

One needs to show there is a polynomial-time algorithm for
decoding whatever code that is utilized in the security proof. If
a non-optimal efficient algorithm is used, its effect on security
needs to be quantified. In this connection, it may be observed
that there is no rigorous quantitative bounds on the efficiency
and security levels of the Cascade protocol \cite{br94} widely
used in experimental implementations. \vspace{3mm}

(2) The Stoping Rule Problem:

As discussed in section IV, to obtain $P(I_E> \delta|pass)$ that
provides quantitative security guarantee of the protocol, one
needs an explicit stopping/ re-starting strategy. Then one needs
to bound Eve's optimal attack performance with respect to such a
strategy from an appropriate overall criterion. \vspace{3mm}

(3) The Future Key Use Problem:

When the generated key $K^g$ is used as one-time pad or as a seed
key in a conventional cipher, a known-plaintext attack on part of
$K^g$ can be combined with Eve's original probe/measurement to
tell something about the rest of the key. This problem arises also
in the KCQ approach, and shows the crucial importance of the
direct encryption known-plaintext attack problem in using keys
obtained form a key-generation protocol.
\section{On criticisms of $\alpha \eta$}\label{app:aecrit}
There have appeared three papers
\cite{lo4,nish,nis2} in quant-ph this year
that purport to show that the $\alpha \eta$ scheme reported in
\cite{barb,corn}
is insecure in various ways. These criticisms are briefly
summarized and responded to in this Appendix.

A general criticism seems to be made in Ref. \cite{lo4}
 that our claim in \cite{barb,corn}
 on the possible use of amplifiers in coherent-state
cryptosystems cannot be valid. It is not clear exactly what this
objection is. In any event, we qualify such use in our papers by
the statement that security must be guaranteed for E attacking near
the transmitter, since quantum amplifiers generally degrade the
communication performance. There are three specific criticisms
from Ref. \cite{lo4}  that one can ascertain:

\vspace{3mm}

(i) In the presence of loss so large that E can get $2^{|K|}$ copies
by splitting the coherent-state signal at the transmitter, there can
only be complexity-based security.

(ii) With just a $3$ dB loss, use of the Grover Search implies
there can only be complexity-based security.

(iii) The Grover Search is `powerful' against complexity-based security.

\vspace{3mm}

In response, observe that only complexity-based security is ever
claimed in \cite{barb,corn} against joint attack on
 the key
$K$. The other information-theoretic security claimed is on
individual ciphertext-only attack on the data. As discussed in
Section VII,
 it is quite sufficient to have complexity-based security
if it can be proved exponential, which is only reduced by a factor of
$2$ with the Grover Search. Long keys of thousands of bits can be
used in $\alpha \eta$ at high speed both in software and hardware
implementation, making the exponential search
completely ineffective.

Furthermore, the Grover Search cannot be launched against
$\alpha \eta$ with a $3$ dB loss. If it can, there is no need for
the $2^{|K|}$ copies extensively discussed in Ref. \cite{lo4}.
Indeed, there is no discussion there on how the Grover or any
search can be launched with a $3$ dB loss. There is a general
misclaim in some papers on quantum cryptography that a $3$ dB
loss on a coherent state cryptosystem renders it insecure because
E can obtain a copy of the quantum ciphertext identical to B. This is
not true even without the use of a secret key.
 It is not true for B92 \cite{ben2} or YK \cite{yuen} or the
usual-state scheme described in \cite{yue5},
 although it renders a coherent-state BB84 scheme
 and some  `continuous -variable' schemes
 essentially insecure. When  a secret key $K$ is used, it is
not true at all. Indeed, the possibility of key generation while granting
E a full copy to bound her performance depends on this being not
true. As explained in this paper, knowledge of the secret key allows
B to make a better measurement than E, who cannot attain the
same performance as B even if she knows the key later.

It is true that when $2^{|K|}$ copies are available to E, there is no
information-theoretic security left in a known-plaintext attack on
direct encryption, and key generation is impossible. But it is clear that
no one would contemplate the operation of cryptosystem over such
a huge loss $2^{-|K|}$ without intermittent amplifiers or other
compensating devices. Numerically, $2^{-|K|}$ corresponds to the
propagation loss over one thousand kilometers of low-loss fibers
without amplifiers for just $|K|\sim 80$ bits in $\alpha \eta$ with
$M\sim 10^3$.
It is totally out of the realm of possibility to sustain such loss even in
ordinary optical communication without cryptography.
As the use of amplifiers is suggested in
\cite{barb,corn}, it is hard to see why
such a criticism is relevant. As a matter of fact, $\alpha \eta$
in its original form is
insecure at a much smaller loss than $2^{-|K|}$ for any reasonable
$|K|$.

In Ref. \cite{nish}, it is claimed that a device can be found that would
 lead
to a bit error rate $P_b^E$ much lower than the quantum detection
theory result $\sim 1/2$ reported in \cite{barb} for individual
ciphertext-only
attack on the data. As pointed out by several others including
G. Barbosa and O. Hirota, such a device cannot exist because
violating quantum detection theory means violating the laws of
quantum physics.
In Ref. \cite{nis2}, it is claimed that $\alpha \eta$ is merely
a classical cipher. The exact nature of  $\alpha \eta$ for key
generation has been analyzed in section V, and for direct encryption
in section VII. While  $\alpha \eta$ can indeed be run in the
classical limit and even in just software, the blanket claim that
 $\alpha \eta$ is classical for intermediate and large signal energy,
 and hence
 presumably does not permit key generation, is incorrect because
 their equation (10) does not hold exactly. In particular, the
 discussion around (\ref{C}) in our section V.A shows how
 $\alpha \eta$ in just its original form may allow key generation.
 For a further concise discussion, see Ref. \cite{yu20}.

\end{document}